\documentclass{nature}

\newcounter{firstbib}

\usepackage{graphicx}
\usepackage{epsfig}
\usepackage[font={small}]{caption}
\usepackage{tablefootnote}

\def\Msun{{\rm M}_\odot}
\def\Lsun{{\rm L}_\odot}
\def\Rsun{{\rm R}_\odot}

\def\Mdot{\dot{M}}
\def\Msunyr{\Msun\,{\rm yr^{-1}}}

\def\gtorder{\mathrel{\raise.3ex\hbox{$>$}\mkern-14mu
             \lower0.6ex\hbox{$\sim$}}}
\def\ltorder{\mathrel{\raise.3ex\hbox{$<$}\mkern-14mu
             \lower0.6ex\hbox{$\sim$}}}

\def\dqy{iPTF\,13dqy}
\def\fs{SN\,2013fs}

\newcommand{\aap}{Astron. Astrophys.}
\newcommand{\araa}{Ann. Rev. Astron. Astrophys.}
\newcommand{\apj}{Astrophys. J.}
\newcommand{\aj}{Astron. J.}
\newcommand{\apjl}{Astrophys. J. Letters}

\newcommand{\nat}{Nature}
\newcommand{\pasp}{Publications of the Astronomical Society of the Pacific}

\newcommand{\mnras}{Mon. Not. R. Astron. Soc.}

\title{Confined Dense Circumstellar Material Surrounding a Regular Type II Supernova:
The Unique Flash-Spectroscopy Event - SN\,2013fs}

\author{O.~Yaron$^{1}$, D.~A.~Perley$^{2,3}$, A.~Gal-Yam$^{1}$, J.~H.~Groh$^{4}$, A.~Horesh$^{5,1}$, E.~O.~Ofek$^{1}$, S.~R.~Kulkarni$^{2}$, 
J.~Sollerman$^{6}$, C.~Fransson$^{6}$, A.~Rubin$^{1}$, P.~Szabo$^{1}$, N.~Sapir$^{1,7}$, F.~Taddia$^{6}$, S.~B.~Cenko$^{8,9}$,
S.~Valenti$^{10}$, I.~Arcavi$^{11,12}$, D.~A.~Howell$^{11,12}$, M.~M.~Kasliwal$^{2}$, P.~M.~Vreeswijk$^{1}$, D.~Khazov$^{1}$,
O.~D.~Fox$^{13}$, Y.~Cao$^{2}$, O.~Gnat$^{5}$, P.~L.~Kelly$^{13}$, P.~E.~Nugent$^{13,14}$, A.~V.~Filippenko$^{13}$, R.~R.~Laher$^{15}$,
P.~R.~Wozniak$^{16}$, W.~H.~Lee$^{17}$, U.~D.~Rebbapragada$^{18}$, K.~Maguire$^{19}$, M.~Sullivan$^{20}$, M.~T.~Soumagnac$^{1}$
}

\begin{document}

\maketitle

\begin{affiliations}
 \item Department of Particle Physics and Astrophysics, Weizmann Institute of Science, Rehovot 76100, Israel.       
 \item Division of Physics, Math and Astronomy, California Institute of Technology, 1200 E. California Boulevard, Pasadena, CA 91125, USA.
 \item Dark Cosmology Centre, Niels Bohr Institute, University of Copenhagen Juliane Maries Vej 30, 2100 Copenhagen \O, Denmark.
 \item School of Physics, Trinity College Dublin, Dublin 2, Ireland.
 \item Racah Institute of Physics, Hebrew University, Jerusalem 91904, Israel.
 \item The Oskar Klein Centre, Department of Astronomy, Stockholm University, AlbaNova, 10691 Stockholm, Sweden.
 \item Plasma Physics Department, Soreq Nuclear Research Center, Yavne 81800, Israel.
 \item Astrophysics Science Division, NASA Goddard Space Flight Center, Mail Code 661, Greenbelt, MD 20771, USA.
 \item Joint Space-Science Institute, University of Maryland, College Park, MD 20742, USA.
 \item Department of Physics, University of California, 1 Shields Ave, Davis, CA 95616-5270, USA
 \item Las Cumbres Observatory, 6740 Cortona Drive, Suite 102,Goleta, CA  93117, USA.
 \item Department of Physics, University of California, Santa Barbara, CA 93106-9530, USA.
 \item Department of Astronomy, University of California, Berkeley, CA 94720-3411, USA.
 \item Computational Research Division, Lawrence Berkeley National Laboratory, 1 Cyclotron Road MS 50B-4206, Berkeley, CA 94720, USA.
 \item Spitzer Science Center, California Institute of Technology, Pasadena, CA 91125, USA.
 \item Los Alamos National Laboratory, Mail Stop B244, Los Alamos, NM 87545, USA.
 \item Instituto de Astronom\'{\i}a, Universidad Nacional Auton\'{o}ma de M\'{e}xico, Apdo. Postal 70-264, 04510 M\'{e}xico DF, M\'{e}xico.
 \item Jet Propulsion Laboratory, California Institute of Technology, Pasadena, CA 91109, USA.
 \item Astrophysics Research Centre, School of Mathematics and Physics, Queens University Belfast, Belfast BT7 1NN, UK.
 \item School of Physics and Astronomy, University of Southampton, Southampton, SO17 1BJ, UK.
 
\end{affiliations}


\newpage

\begin{abstract}
With the advent of new wide-field, high-cadence optical transient surveys, our understanding of the diversity of
core-collapse supernovae has grown tremendously in the last decade. However, the pre-supernova evolution of massive
stars, that sets the physical backdrop to these violent events, is theoretically not well understood and difficult to
probe observationally. Here we report the discovery of the supernova \dqy\ = \fs\, a mere $\sim3$ hr after explosion.
Our rapid follow-up observations, which include multiwavelength photometry and extremely early (beginning at $\sim6$ hr
post-explosion) spectra, map the distribution of material in the immediate environment ($\ltorder 10^{15}$\,cm) of the
exploding star and establish that it was surrounded by circumstellar  material (CSM) that was ejected during the final
$\sim1$\,yr prior to explosion at a high rate, around $10^{-3}$ solar masses per year. The complete disappearance of
flash-ionised emission lines within the first several days requires that the dense CSM be confined to within $\ltorder
10^{15}$\,cm, consistent with radio non-detections at 70--100 days. The observations indicate that \dqy\ was a regular
Type II SN; thus, the finding that the probable red supergiant (RSG) progenitor of this common explosion ejected material
at a highly elevated rate just prior to its demise suggests that pre-supernova instabilities may be common among exploding
massive stars.
\end{abstract}


Why and how massive stars explode as supernovae is one of the outstanding open questions in astrophysics. Massive stars
fuse light elements into heavier ones in their core. During the final years of their (relatively short, a few
$10^6$--$10^7$ yr) lifetime, these stars burn heavy fuel, the fusion products of hydrogen and helium, until an iron core
grows and ultimately collapses. Stellar evolution in these final years, which sets the initial conditions for the
final collapse and explosion of such stars as supernovae (SNe), is poorly understood\cite{2012ARA&A..50..107L}. Direct
observations of these processes is challenging, as stars in these brief final stages are rare. Statistically, it is very
likely that not even a single star that is within 1\,yr of explosion currently exists in our Galaxy.

Recently, growing observational evidence has suggested the existence of pre-explosion
elevated mass loss and eruptions\cite{2014Natur.509..471G, 2014ApJ...789..104O, 2013MNRAS.430.1801M,
2008MNRAS.389..113P}. Accommodating these findings, a handful of theoretical studies\cite{2015ApJ...810...34W,
2014ApJ...780...96S, 2011ApJ...733...78A, 2010ApJ...717L..62Y} were carried out exploring possible pathways by
which massive stars may become unstable during their terminal years, leading to the observable signatures of
increased mass loss, variability, and eruptive episodes prior to the terminal explosion. 
Material ejected by the star in the year prior to its demise may imprint unique signatures on the emission observed from the young 
SN event, but as this material will be quickly swept away by the expanding explosion debris, such detections require rapid
observations to be secured within a few days of explosion\cite{2014Natur.509..471G, 2016ApJ...818....3K}. 
A handful of recent observations provide evidence for enhanced mass loss and eruptive
episodes during the terminal years prior to explosion, but mainly for rare subclasses of SNe which comprise at
most a few percent of the population. 
The observations presented here of \dqy\ indicate that it was a fairly regular Type II SN, similar to $\sim50\%$\cite{2011MNRAS.412.1522S} of exploding massive stars,
and thus may strengthen the hypothesis that the ultimate collapse of the core and the
preceding vigorous ejection of mass from the outer envelope are causally coupled.
In addition, the structure of the outer envelope of massive stars during the very late stages of evolution may
significantly differ from what is predicted by stellar evolution models\cite{2013A&A...559A..69G, 2006MNRAS.367..186E}.

On 2013 Oct. 6.245 (UTC dates are used throughout this paper), a new transient source with 
$r_{\rm PTF}=18.6 \pm 0.05$\,mag was identified within the nearby galaxy
NGC\,7610 (redshift $z=0.011855$ (NED), $d=50.95$\,Mpc) by the automated real-time discovery and classification pipeline 
of the intermediate Palomar Transient Factory (iPTF) survey\cite{2009PASP..121.1395L}. 
A second image, confirming the detection, was obtained 50\,min later on Oct. 6.279 (Supplementary Fig.~\ref{SIfig-discovery}).
The transient was automatically saved in the survey's database on Oct. 6.365 and was internally designated as \dqy.
This event was independently discovered by K. Itagaki a day later on Oct. 7.468 and was  
assigned the name \fs\cite{2013CBET.3671....1N}.

\begin{figure}
\centering
\includegraphics[width=16cm]{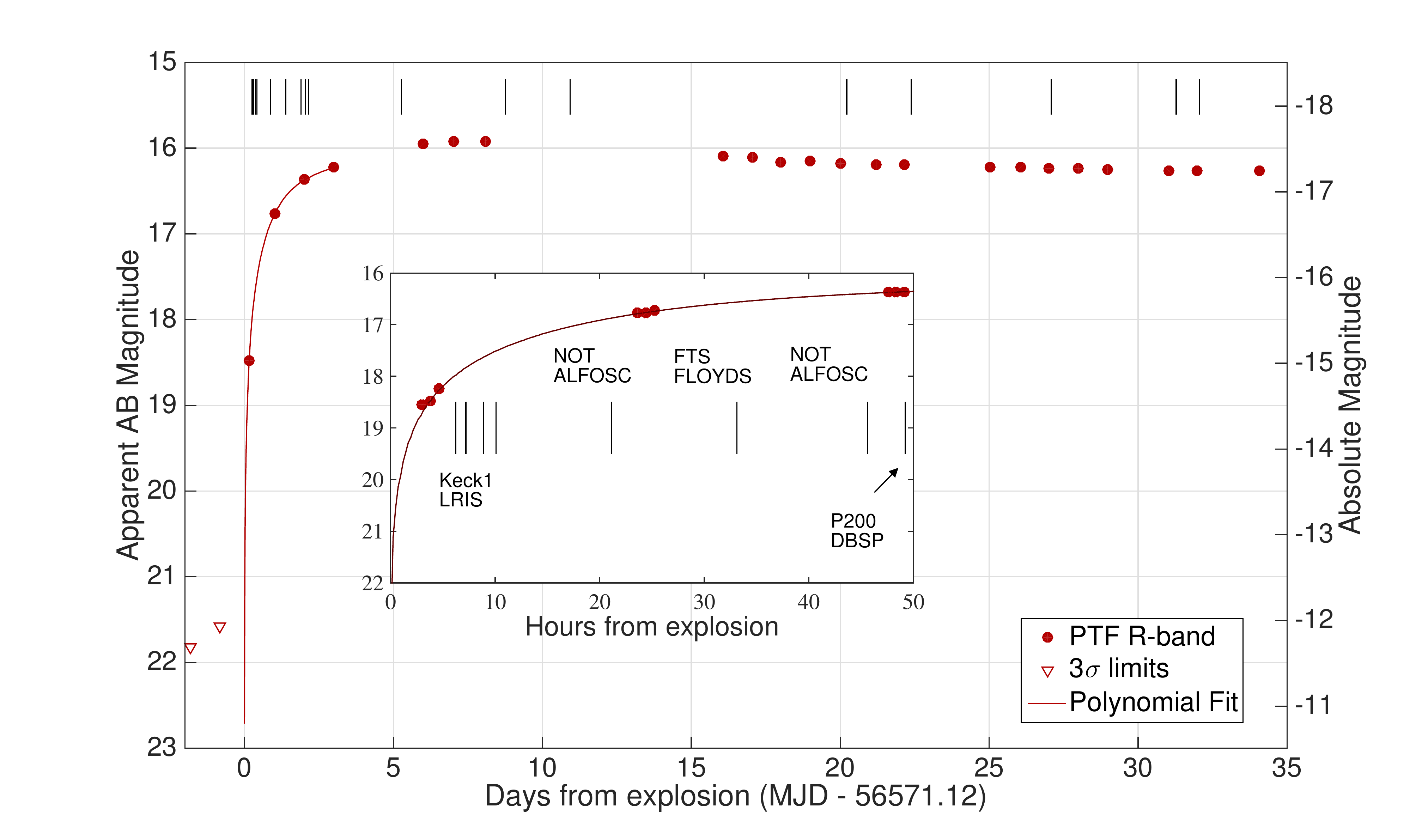}
\includegraphics[width=16cm]{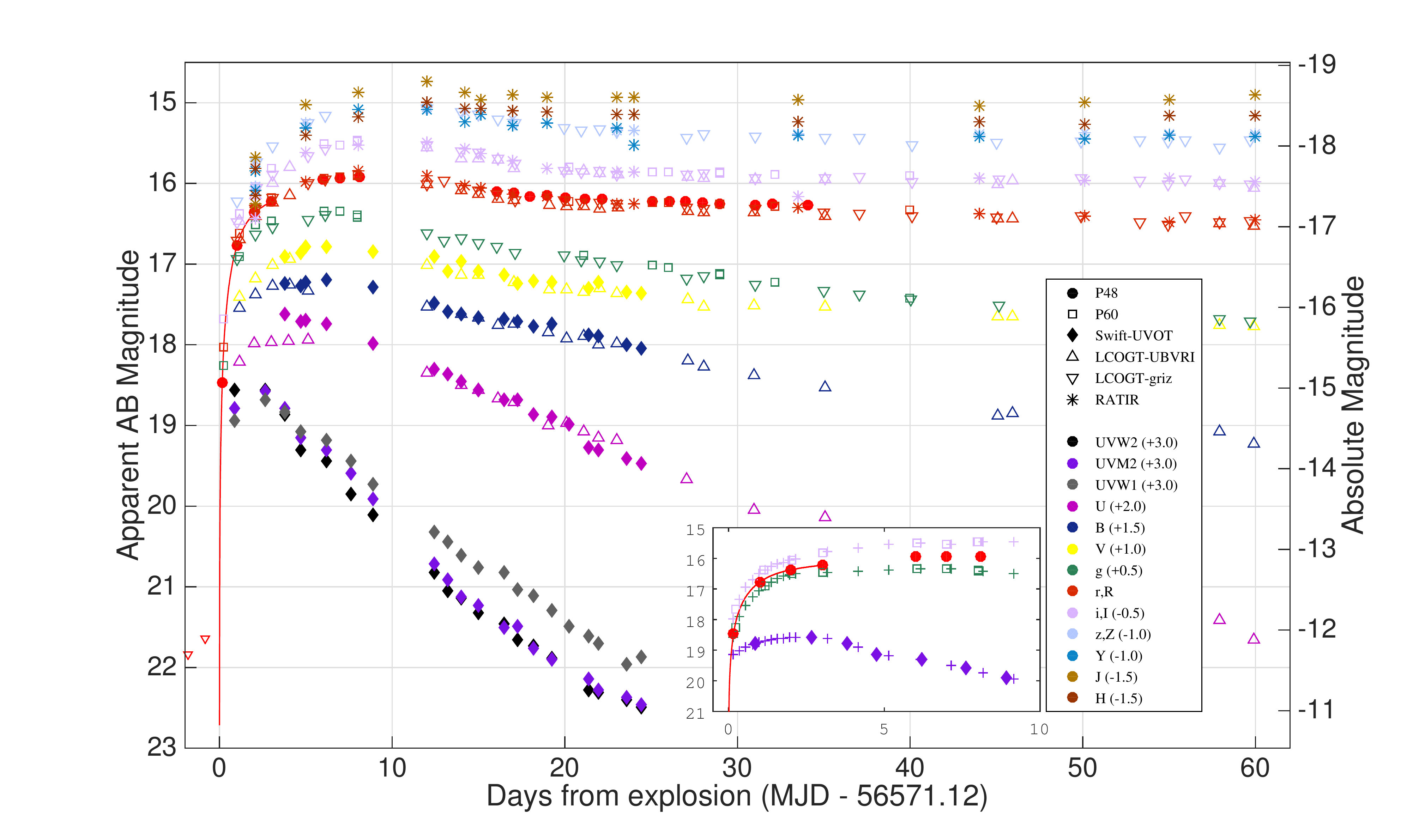}
\caption{The early discovery of \dqy\ and the prompt ensuing follow-up observations, multiband photometry and spectroscopy, 
enabled the direct probing and mapping of the nearby environment surrounding the progenitor.
Top: A polynomial fit to the Palomar 48-inch $r$-band light curve is used to estimate that the explosion occurred on 2013 Oct. 6.12;
thus, this event was caught $\sim3$\,hr after explosion and the first spectrum was obtained $\sim6$\,hr post-explosion.  
The $r$-band light curve attains a peak absolute magnitude of $M_R = -17.40 \pm 0.15$ and settles on a plateau at $M_R = -17.00 \pm 0.15$\,mag.
Pre-explosion $3\sigma$ upper limits (nondetections) are denoted with triangles.
Dates of spectroscopic observations are marked with vertical lines at the top. See inset for the first 50\,hr (with unbinned photometry points).
Bottom: Multicolour UVOIR observations, alongside the observed spectral evolution (Fig.~\ref{fig-latespec}), indicate a resemblance to a Type II-P core-collapse SN.
The various bands have been shifted in magnitude with respect to the $r$ band for clarity, as noted in the legend.
The inset shows a backward extrapolation of selected light curves to the date of discovery (the first P48 detection; see methods for details).
\label{fig-photometry}
}
\end{figure}

\begin{figure}
\centering
\includegraphics[width=17cm]{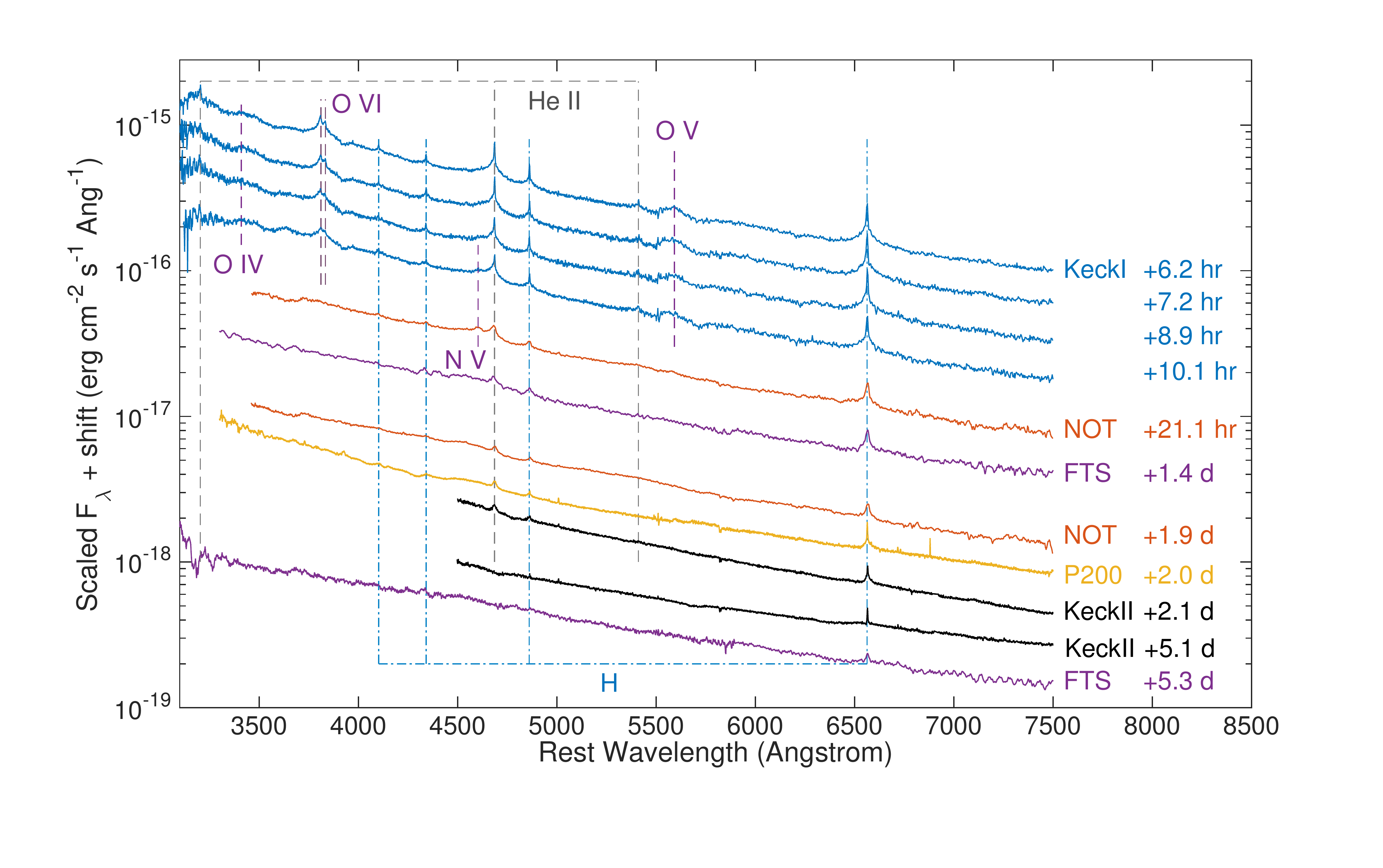}
\includegraphics[width=17cm]{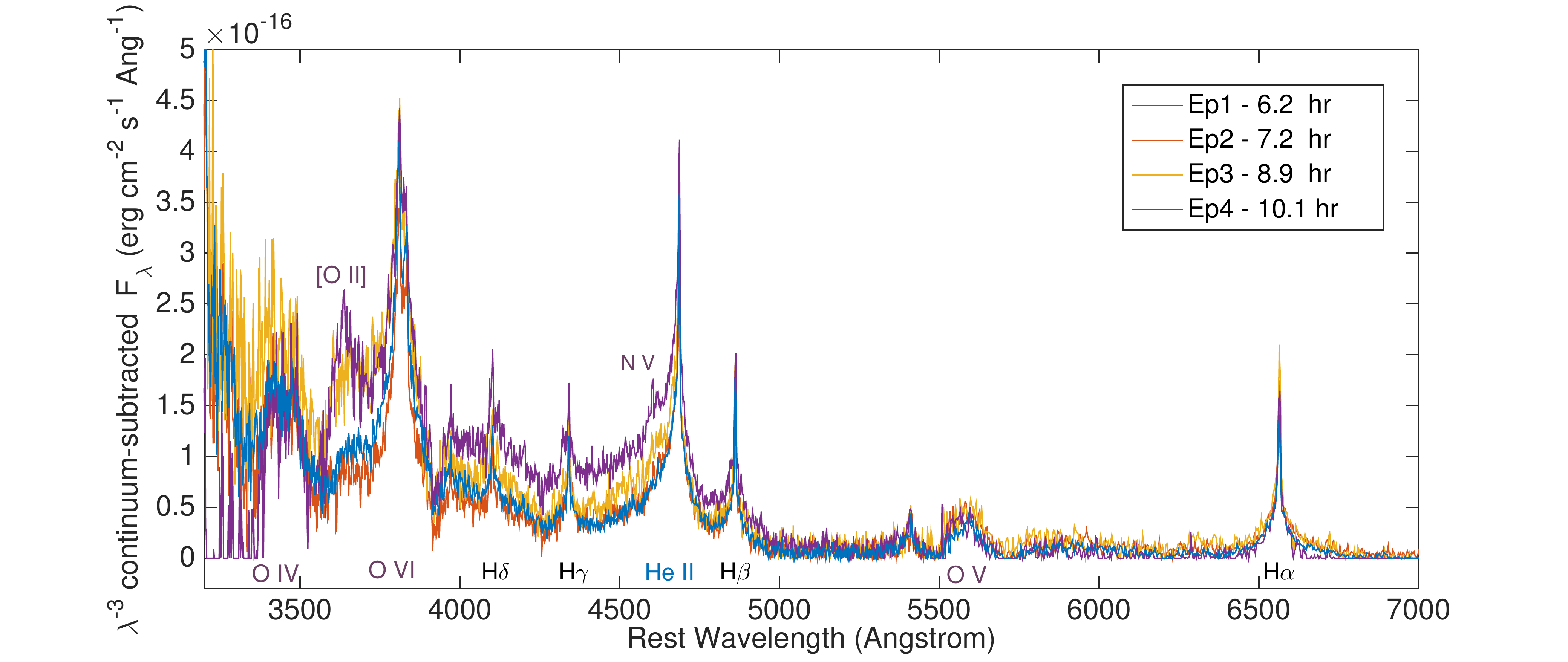}
\caption{Early-time flash spectroscopy of \dqy\ reveals flash-ionisation signatures during the first 2\,d after explosion. 
High-ionisation oxygen emission lines (O~VI $\lambda\lambda$3811, 3834, O~V $\lambda5597$, O~IV $\lambda3410$) dominate during 
the first 6--10\,hr (the O~V feature is affected by an instrumental artifact; see Methods \S\ref{sec:specmodel}), 
while N~V $\lambda4604$ and He~II $\lambda4686$ persist for 21\,hr and 2.1\,d, respectively, along with the hydrogen Balmer series.
All throughout these first several days the underlying continua are blue, and by day 5 the spectrum turns nearly featureless. 
The indicated epochs are all with respect to our adopted explosion time (Fig.~\ref{fig-photometry}).
An overlay of the four early-time Keck spectra (6--10\,hr), after subtraction of a 
$\lambda^{-3}$ curve, normalised to fit the continuum of each spectrum in the wavelength range 5000--6000\,\AA, 
is shown in the bottom panel, revealing the similarity of the emission features during this 4\,hr timespan.
N~V begins to emerge in the last epochs (9--10\,hr).
All spectra have been calibrated to the PTF $r$-band photometry and corrected for redshift and Galactic extinction ($E(B-V)_{\rm MW}=0.035$\,mag).
\label{fig-earlyspec}}
\end{figure}

\begin{figure}
\centering
\includegraphics[width=17cm]{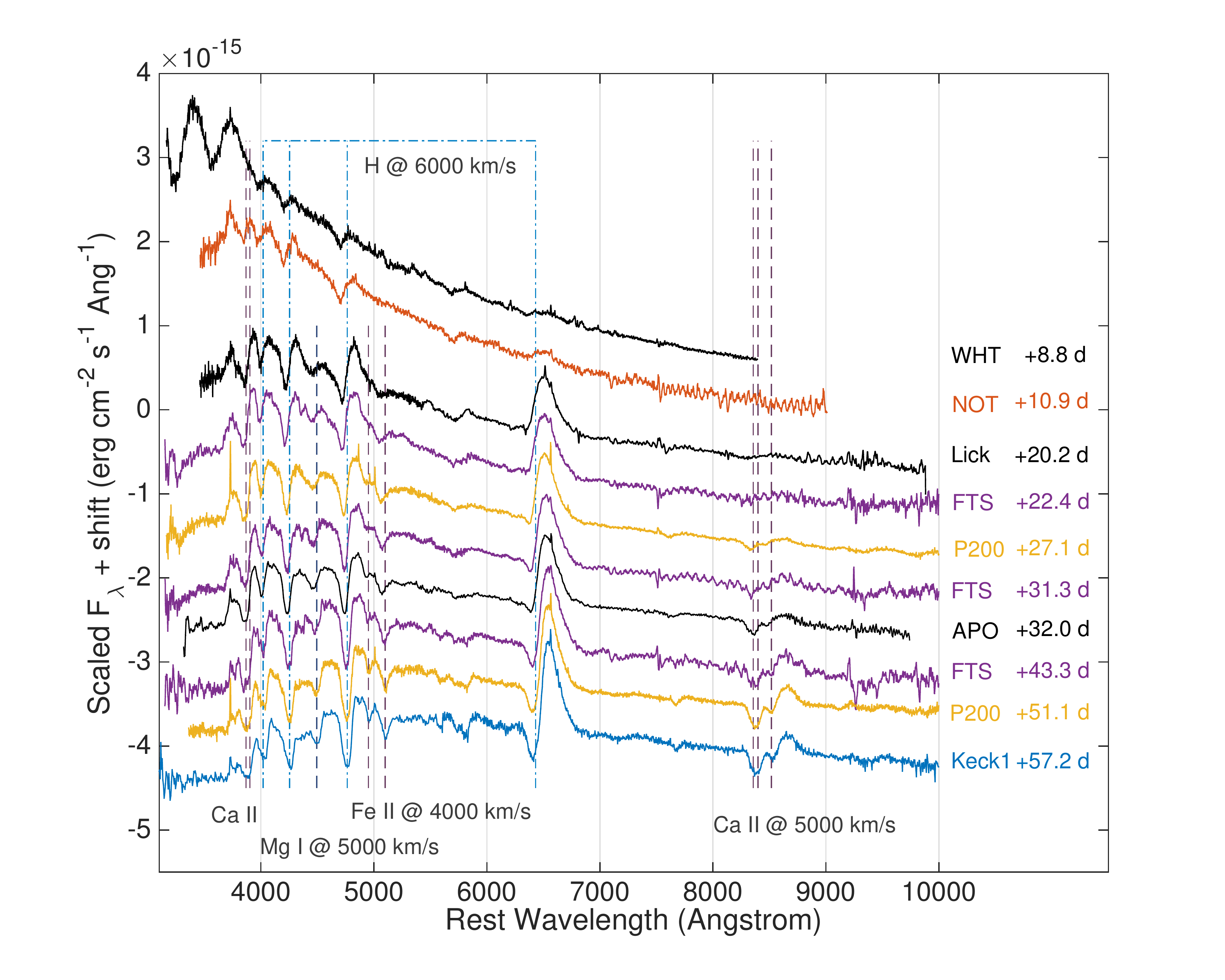}
\caption{Spectra of \dqy\ obtained between 8 and 57 days after explosion, 
showing P-Cygni profiles of the Balmer series and additional typical elements,
confirm a regular Type II SN identification.
The Balmer series, Ca~II (the near-infrared triplet and the H\&K lines), 
Fe~II, and Mg~I are marked 
for the specified expansion velocities.
See the spectroscopic log (Table~\ref{SItab-spec}) for details of the spectra.
All spectra presented in this study are publicly available via WISeREP\cite{2012PASP..124..668Y} (http://wiserep.weizmann.ac.il).
\label{fig-latespec}}
\end{figure}

Follow-up observations were promptly initiated, 
including X-ray and multicolour ultraviolet-optical-infrared (UVOIR) photometry (Fig.~\ref{fig-photometry}) and 
multi-epoch rapid spectroscopy (Fig.~\ref{fig-earlyspec}).
A third-order polynomial fit to our early flux measurements (Fig.~\ref{fig-photometry}, top, inset) is used to estimate that the first light 
(shock breakout following the explosion) occurred on Oct. 6.12 ($\pm0.02$\,d), $\sim3\pm0.5$\,hr before the first detection. 
A series of spectra, the earliest ever taken of a SN, were obtained using LRIS mounted on 
the 10\,m Keck-I telescope to follow the evolution of flash-ionised emission lines (Fig.~\ref{fig-earlyspec}).
Our flash spectroscopy\cite{2014Natur.509..471G} sequence initially (6--10\,hr post-explosion) shows unprecedented strong 
high-ionisation emission lines of oxygen (O~IV, O~V, O~VI), which disappear within the following 11\,hr.
Our $t=21$\,hr spectrum still shows He~II and N~V; the He~II lines persist until day 2.1. 
Throughout this period, hydrogen Balmer lines of decreasing strength are observed. 
Unlike SN\,2013cu (iPTF13ast)\cite{2014Natur.509..471G} and SN\,1998S\cite{2015ApJ...806..213S},
which show prominent ionised nitrogen lines in their early-time spectra, nitrogen lines in the early spectra of \dqy\ 
are much weaker than those of oxygen.
The width of the narrow core of the H$\alpha$ line measured from our highest-resolution spectrum 
(see Sec. \ref{sec:spec} of the Methods; hereafter Methods \S\ref{sec:spec}; Supplementary Fig.~\ref{SIfig-halpha_hires}) 
indicates an upper limit on the expansion velocity of the emitting material of $100\,{\rm km\,s^{-1}}$.
By day 5 the spectra remain blue, but no longer show prominent emission features.
Later spectra (see Methods \S\ref{sec:spec}, Fig.~\ref{fig-latespec}) develop broad P-Cygni lines typical of Type~II SNe.
Along with the flat, plateau $r$-band light-curve evolution (Fig.~\ref{fig-photometry}, bottom), our observations indicate that \dqy\ resembles a Type II-P SN. 
The evolution of the expansion velocity of the SN ejecta for the ${\rm H\alpha, H\beta}$, and Fe~II $\lambda 5169$ lines (shown in Supplementary Fig.~\ref{SIfig-velevol}) is also in agreement with the behaviour of regular SNe~II\cite{2006ApJ...645..841N, 2014MNRAS.442..844F, 2010MNRAS.404..981M}. 

Following ref.~2, 
we interpret the early spectral evolution as the result of flash ionisation and recombination of dense CSM surrounding the progenitor of \dqy. 
Using the methodology and model assumptions similar to those of ref.~20 
(see Methods \S\ref{sec:specmodel}), we produced model spectra for varying combinations of  the inner boundary (emitting) radius, $R_{\rm in}$, 
the bolometric luminosity at this inner boundary, $L_{\rm SN}$, and the mass-loss rate from the progenitor star, $\Mdot$.
All models were produced for abundances that are consistent with solar, allowing for enhancement of the surface helium abundance, 
and a progenitor wind velocity of $v_{\rm wind}=100\,{\rm km\,s^{-1}}$. 
We find values of $\Mdot=(2$--4) $\times 10^{-3}\,\Msunyr$, $R_{\rm in}=(1.3$--1.4) $\times 10^{14}\,{\rm cm}$, $L_{\rm SN}=(2.0$--3.5) $\times 10^{10}\,{\rm \Lsun}$, and associated effective temperatures ($T_{\rm in}$, according to the Stefan-Boltzmann law) at the base of the wind
in the range $\sim48$--58\,kK.

\begin{figure}
\centering
\includegraphics[width=14cm]{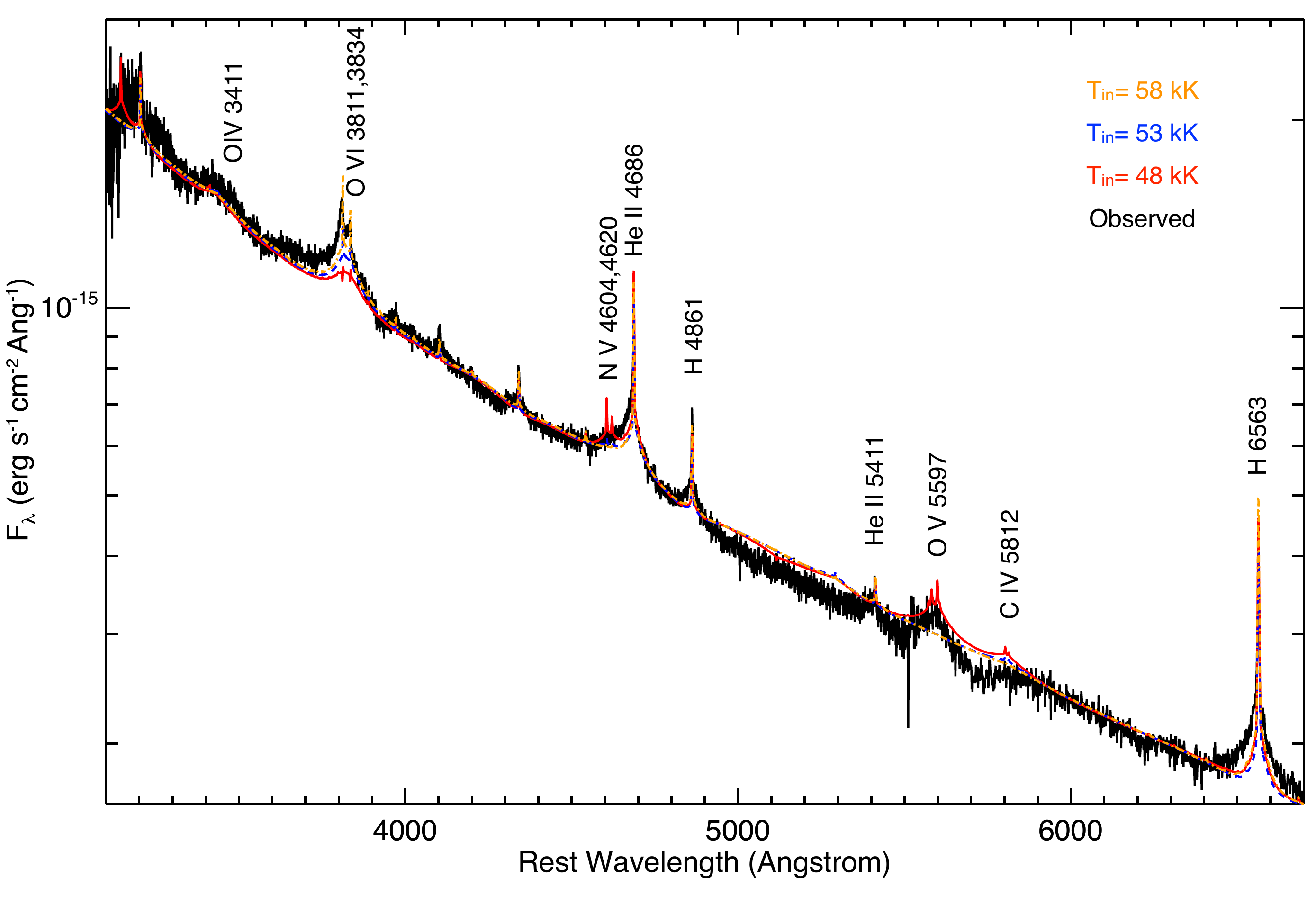}
\includegraphics[width=14cm]{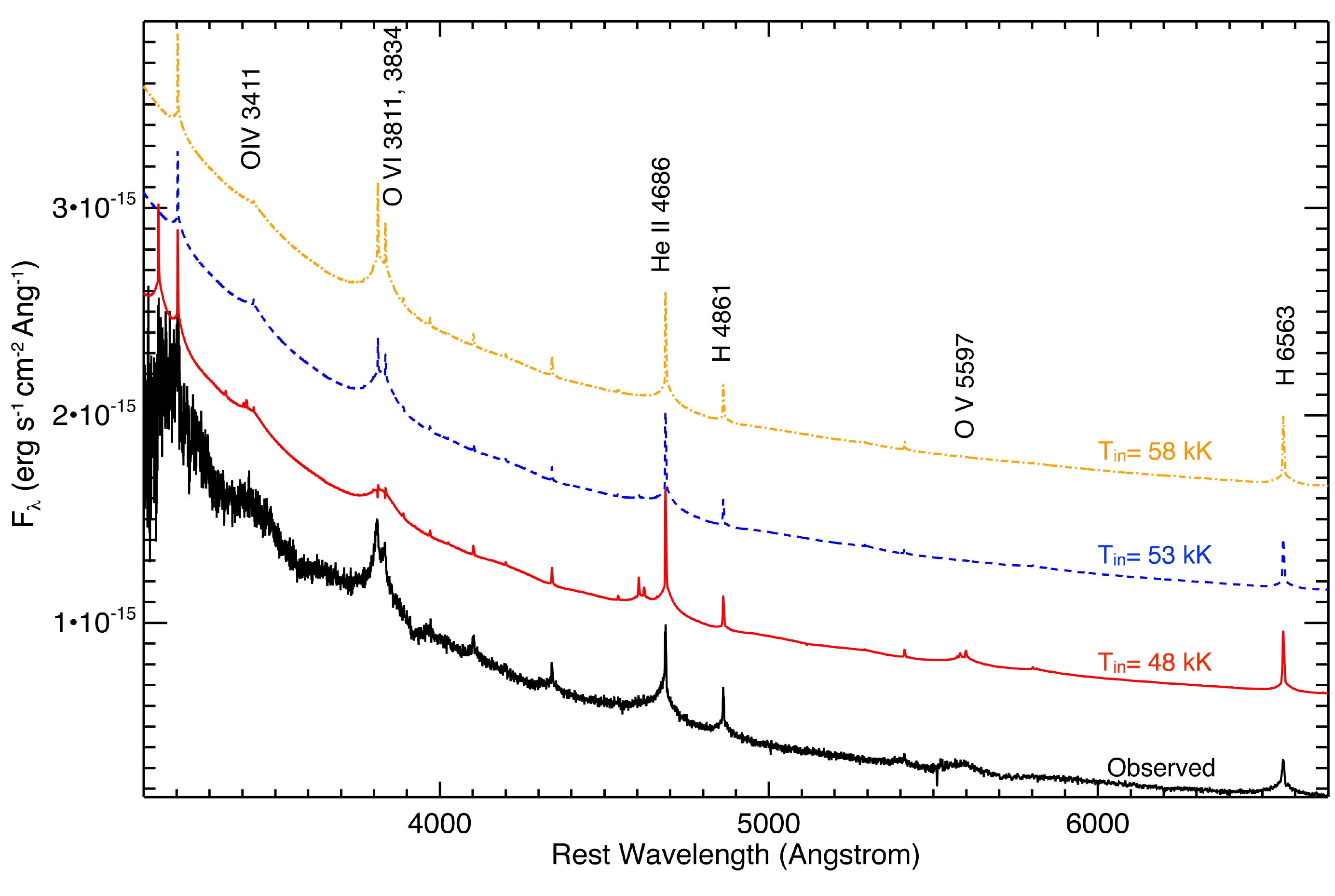}
\caption{The best-fitting CMFGEN models to the observed early-time spectrum (6\,hr post-explosion) of \dqy\ reveal a temperature
of the line forming region around $50$--$60$\,kK.
The model spectra are shown in colour (top: all models over-plotted, bottom: shifted for clarity), the early Keck spectrum is shown in black 
(applying a total reddening of $E(B-V)_{\rm tot}=0.05$\,mag and $R_V=3.1$ to best match the continuum spectral energy distribution).
Line features (bottom) indicate that the $58$\,kK model (top dot-dashed orange curve) does not recover the O~V line, while a cooler model ($48$\,kK; red) predicts too weak O~VI lines.
The three model spectra bracket the observed early-time spectrum; the $53$\,kK model (dashed blue) best recovers the oxygen ionisation structure.
See Methods \S\ref{sec:specmodel} for details.
\label{fig-Groh_spec}}
\end{figure}

Fig.~\ref{fig-Groh_spec} displays the comparison of the obtained model spectra to the first Keck spectrum at $\sim6$\,hr after the explosion.
As evident from the plots, the models bracket the observed early spectrum.
The model for the highest $T_{\rm in}=58.5$\,kK has virtually no O~V or O~IV emission, 
while the 53\,kK model does match the observed features well and seems to follow best the overall ionisation structure.
This model should be considered illustrative only, as it is possible that the temperature structure of the outflow is not fully reproduced by the model (e.g., due to a nonspherical geometry or a light-travel-time effect).
However, we can estimate that the average effective temperature at the base of the outflow is $T_{\rm in}\approx53\pm5$\,kK,
and also that the abundances of the emitting material are consistent with 
solar composition and increased surface He abundance typical of evolved massive stars.

Since the value of $v_{\rm CSM}=100\,{\rm km\,s^{-1}}$ sets an upper limit on the progenitor wind speed from our observations, 
and typical RSG winds have lower velocities $\sim 10$--15\,km\,s$^{-1}$ (e.g., ref.~21), 
we also calculated models for $v_{\rm CSM}=15\,{\rm km\,s^{-1}}$.
Since $\Mdot$ scales linearly with $v_{\rm wind}$ (e.g., Eq.~1 of ref.~20, 
these models resulted in almost an order of magnitude decrease in the mass-loss-rate estimate, 
$\Mdot=(3$--6) $\times 10^{-4}\,\Msunyr$,
with $R_{\rm in}$ and  $L_{\rm SN}$ almost identical to the above values, obtained for the higher $v_{\rm CSM}$ (see Methods \S\ref{sec:specmodel}).
We therefore derive an overall mass-loss rate of $\Mdot=(0.3$--4) $\times 10^{-3}\,\Msunyr$
for the possible range of $v_{\rm CSM} \approx 15$--100\,km\,s$^{-1}$.
However, it should be emphasised that although typical RSG wind velocities are well below $\sim100\,{\rm km\,s^{-1}}$, 
a short, elevated, perhaps eruptive mass-loss episode may result from 
a different physical mechanism than that driving a normal wind, and thus have higher wind velocities.
On top of that, the O~VI doublet ($\lambda\lambda$3811, 3834) lines that are resolved in our earliest spectra 
show Doppler broadening corresponding to velocities significantly higher than  $100\,{\rm km\,s^{-1}}$. 
This line is formed in the inner parts of the outflow, and even if the velocity decreases at larger radii, 
a significant fraction of the outflow (progenitor wind) will be moving with much higher velocities than those assumed for typical quiescent RSGs.

Additional independent lower-limit estimates on the mass-loss rate 
are obtained from the H$\alpha$ luminosity (assuming an $r^{-2}$ spherical wind-density profile),
as well as from the prominent electron-scattering wings of the emission-line profiles during the first several days.
The derived lower limits all point to $\Mdot \gtorder 10^{-3}\,\Msunyr$ (see Methods \S\ref{sec:flux}).

Our models constrain the temperature using the line features and predict a continuum
shape that follows a modified blackbody curve. 
The models can accommodate only a low extinction by dust ($E(B-V)_{\rm tot}=0.05$\,mag)
while maintaining this match. In view of the estimated
Galactic (Milky Way) extinction toward this direction ($E(B-V)_{\rm MW}=0.035$\,mag), the models
indicate a very small residual extinction contributed by the SN host galaxy ($E(B-V)_{\rm host}\approx0.015$\,mag). 
This is consistent with the blue colours
of the star-forming knot at the location of the SN (External Data Fig.~\ref{SIfig-discovery}) 
as well as the nondetection of interstellar material (ISM)
absorption lines (e.g., Na~I~D) from the host. The dearth of local dust
surrounding the SN further supports the lack of {\it persistent} high mass loss from the progenitor, as such
massive outflows are predicted to be locations of efficient dust formation\cite{2013EAS....60..175C}.

Assuming that the material around the progenitor star expanded at $\ltorder 100\,{\rm km\,s^{-1}}$ and 
was swept up by the SN ejecta moving at typical velocities of $10^{4}$\,km\,s$^{-1}$
within $\sim5$\,d after explosion (the spectra of day 5 being blue and featureless), 
the flash-ionised CSM was emitted within $\sim500$\,d (this value increases inversely with the actual wind velocity)
prior to explosion and is confined to within $<5 \times 10^{14}$\,cm, consistent with our modeling results.
Assuming a constant mass-loss rate (for the assumed upper $v_{\rm wind}=100\,{\rm km\,s^{-1}}$) of $\sim 3 \times 10^{-3}\,\Msunyr$, 
as deduced above, the estimate of the total mass that was lost during this pre-explosion 
enhanced mass-loss phase is a few $10^{-3}\,\Msun$, regardless of  $v_{\rm wind}$, at most.

Additional constraints on the physical scale of the emitting region may be derived from the early emission-line evolution.
As seen in Fig.~\ref{fig-earlyspec}, in the spectrum at $21$\,hr, the O~VI emission lines have completely disappeared.
Assuming a short ionising burst from a source surrounded by a spherical shell of material of radius $r$ that promptly recombines, 
the resulting recombination lines will be visible for at least the light-travel time between the point facing the observer 
and points on the equatorial ring perpendicular to the line of sight -- i.e., $r/c$, where $c$ is the speed of light. 
The nondetection of the high-ionisation O lines $21$\,hr after explosion thus places  
an upper limit on the radius of the emitting region, $r \ltorder 2 \times 10^{15}\,{\rm cm}$ (21\,light-hr).
By the same arguments, the persistence of these lines with little evolution over the 4\,hr around 6--10\,hr after explosion 
(Fig.~\ref{fig-earlyspec}, bottom panel) implies a lower limit of $\sim4$\,light-hr, $r \gtorder 4 \times 10^{14}$\,cm. 
The high wind densities that persist throughout the emitting region ($n_e\approx 10^{10}\,{\rm cm^{-3}}$) imply short recombination times, on the order of minutes\cite{2014Natur.509..471G}, so the emitted spectra are expected to respond promptly to the evolving radiation field as assumed.

A possible caveat to the lower limit is that the assumption of a short ionising burst operating for less than 4\,hr may not be valid. 
However, light-travel effects naturally explain the fact that there is hardly any change in the ionisation structure during at least those 4\,hr, 
whereas the model temperature drops significantly (by a factor of $\sim1.5$; Supplementary Fig.~\ref{SIfig-BBTempRad}) 
during this period with little observable spectral evolution.
Our data and models are therefore all consistent with a confined zone of dense CSM 
deposited by the exploding star within a few hundred days prior to its terminal explosion
(compatible with, for example, the proposed delay times from an early hydrogen-rich envelope ejection 
following the violent off-centre silicon degenerate flash and the terminal iron-core collapse
as computed in ref.~6 
for stars with initial masses around $10\,\Msun$).

\begin{figure}
\centering
\includegraphics[width=15cm]{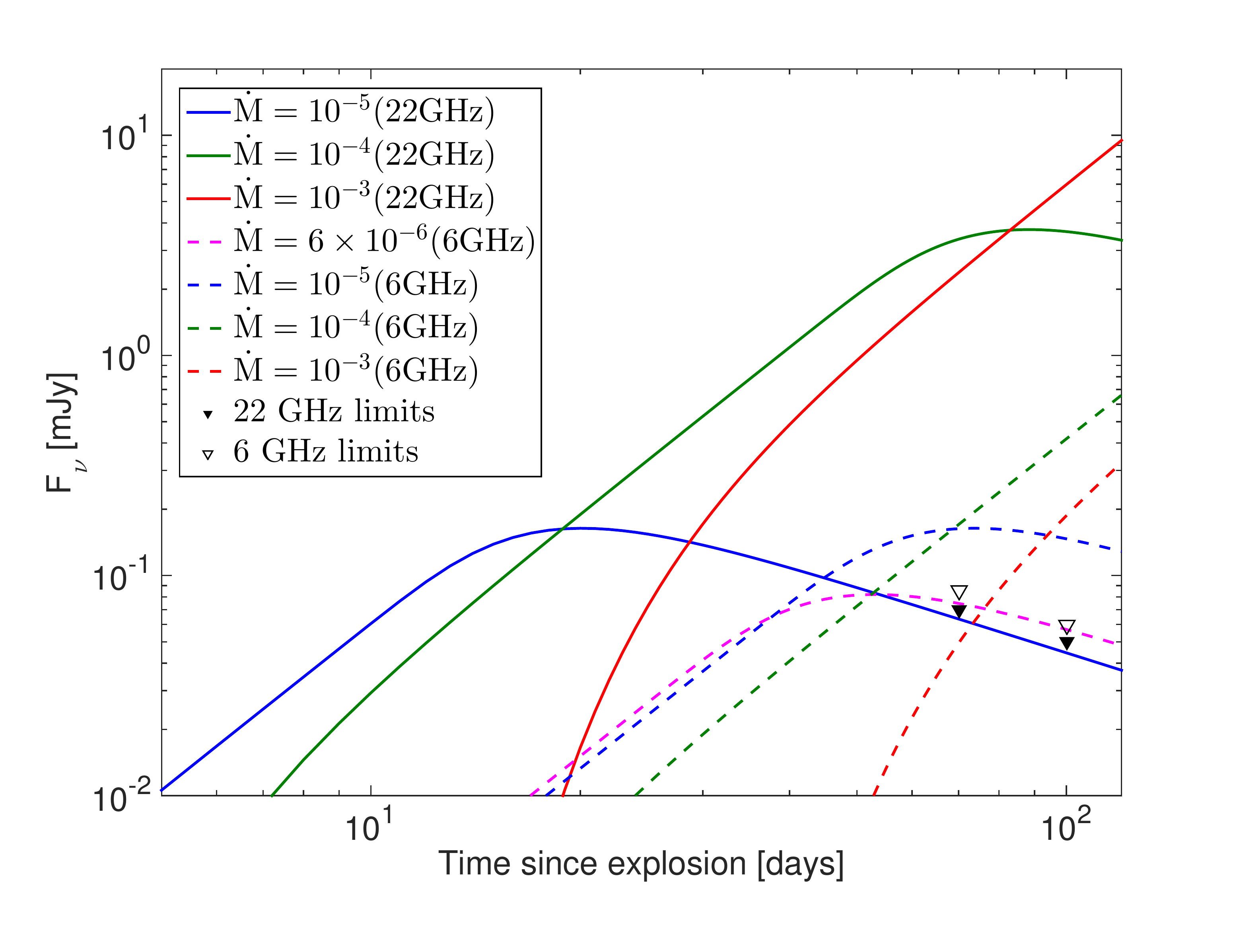}
\caption{Radio non-detections rule out an {\em extended} dense wind structure around the progenitor of \dqy. 
The plotted coloured curves display theoretical light curves of radio emission originating from the interaction of SN ejecta with an extended CSM. These extended CSM structures are assumed to be a result of a constant mass-loss rate through stellar winds. Our measured limits (triangles) on the late-time radio emission (at $\sim70,100$\,d) rule out a wide range of mass-loss rates. In particular, our limits show that the mass-loss rate estimate required to explain the early optical data could not have been sustained over long periods. Otherwise, an extended CSM structure would have formed with sufficiently high density to be detected in the radio at late times
(up to a limiting free-free absorption regime; see Methods \S\ref{sec:radio}). 
Models were calculated assuming wind velocities of $100$\,km\,s$^{-1}$, a SN shockwave expanding at $10^4$\,km\,s$^{-1}$,
and an electron power-law energy distribution with a power-law index of $p=3$ and equipartition.
\label{fig-radio_LCs}
}
\end{figure}

To test whether the CSM with which the SN interacts at early times (and that we detect in the optical spectra) 
is part of an extended CSM structure originating from continuous stellar winds, 
we first consider the lack of narrow lines in spectra obtained after day 5.
Intense mass loss ($\Mdot>10^{-4}\,{\rm M_{\odot}\,yr^{-1}}$) is typical of SNe~IIn\cite{2012ApJ...744...10K} 
and manifests itself in sustained optical emission lines that are not seen in our observations (Fig.~\ref{fig-latespec}).
This argues against a constant, high wind-like mass loss from the progenitor. 
Next, adopting the density estimates we derived from the optical flash spectra, 
we extrapolate the densities to larger distances assuming a wind-like $r^{-2}$ density structure. 
As can be seen in Fig.~\ref{fig-radio_LCs}, such an extended wind would have led to detectable radio emission at cm wavelengths,
for mass-loss rates in the range $6\times 10^{-6} \ltorder \Mdot \ltorder 10^{-3}\,{\rm M_{\odot}\,yr^{-1}}$ and the assumed shockwave and wind velocities.
However,  our Jansky Very Large Array (VLA) observations (see Methods \S\ref{sec:radio}) 
at central frequencies of 6.1\,GHz and 22\,GHz obtained on 2013 Dec. 17, 
about 70\,d after explosion, resulted in null detections (Fig.~\ref{fig-radio_LCs}), as did a second VLA observation obtained a month later.
We consider the possible complicating effects of free-free absorption in Methods \S\ref{sec:radio}.
These strengthen the suggestion for the episodic nature of the elevated pre-explosion mass loss.

X-ray observations, conducted with the {\it Swift} X-Ray Telescope (XRT) between days 1 and 25 after explosion,
set a combined upper limit on the X-ray luminosity of $L_{\rm X} \ltorder 4.7\times 10^{40}\,{\rm erg\,s^{-1}}$ (see Methods \S\ref{sec:phot}).
Preliminary calculations done using customised multigroup hydrodynamic simulations of shock breakout (SBO) in an optically thick, 
steady mass-loss wind, calibrated to the density assumed above (see Methods \S\ref{sec:sbo} for details),
suggest an X-ray luminosity in excess of $10^{42}$--$10^{43}\,{\rm erg\,s^{-1}}$, and thus
reveal that it is likely that the SBO did not take place within the engulfing dense wind.
It is therefore highly plausible that the SBO occurred at the edge of the progenitor star, or within an optically thin CSM ($\tau \approx$ a few).
This further suggests that the dense nearby CSM may have been detached from the surface of the progenitor star.
A full three-dimensional model including SBO and radiative transfer in nonspherical configurations
may be required to fully explore the set of constraints provided by the data.

In view of the above, we model the early-time multiband UVOIR light curves using the methods of ref.~24 
(hereafter, RW11), applicable to the spherical expansion phase following SBO.
We fit an RSG model (without wind) to the multicolour data (Supplementary Fig.~\ref{SIfig-RW11_LCs}) and obtain an estimate of the radius of the exploding star and the energy per unit mass of the explosion (Supplementary Fig.~\ref{SIfig-chi_RW11}). 
The energy value obtained ($\sim5 \times 10^{50}$\,erg, assuming an ejecta mass of $10\,\Msun$) 
is broadly consistent with those expected from previous studies of SNe~II-P (see also ref.~25), 
while the range of progenitor radii (100--350\,$\Rsun; 2\sigma$) is on the low side of generally quoted values, 
though some studies do call for a reduction of RSG radii for typical SNe~II (e.g., ref.~26). 
The models of RW11 
assume an RSG progenitor with a standard density structure. 
However, the mass-loss episode reported here 
(and additional possible mass-loss/stripping episodes during the progenitor's advanced evolutionary stages; see, e.g., 
the study by ref.~27) 
may well have modified the progenitor structure, 
and the escaping radiation may also be altered by reprocessing in the detached CSM shell.
A full theoretical investigation of this physical setup lies beyond the scope of this article.

Ref.~28 
presents an alternative scenario to produce confined CSM around RSG SN progenitors 
due to photoionisation by an external radiation field.
This model, which could potentially remove the need for an elevated mass-loss episode promptly before the SN explosion,
seems to explain static CSM shells more massive than we find here and also at greater radial distances
($M_{\rm shell} \approx 10^{-2}$--$10^{1}\,\Msun,\ R \approx 10^{16}$--$10^{18}$\,cm, 
as specified for typical wind properties and external radiation fluxes).
Additional careful modeling is required to test whether such a scenario, or variants of it, can also explain our observations.

\begin{figure}
\centering
\includegraphics[width=18cm]{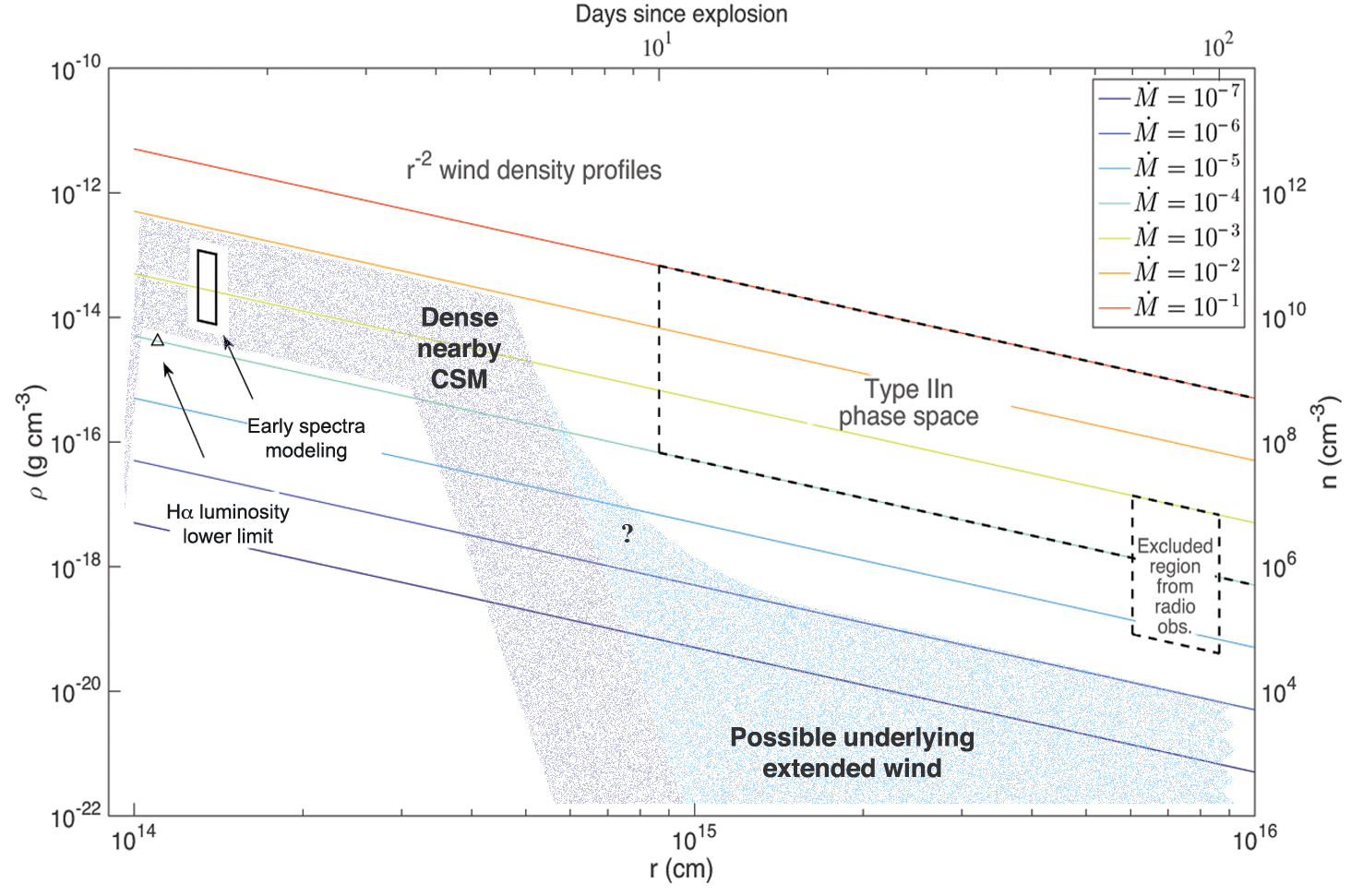}
\caption{The proposed CSM configuration surrounding the progenitor of \dqy, resulting from our multiwavelength observations --- 
a CSM density profile that is both nearby and confined.
The figure summarises the CSM estimates we derive from analyses of the early-time spectra (at $r \approx 10^{14}\,{\rm cm}$) 
together with the region excluded by the later radio observations.
The solid diagonal coloured lines denote constant mass-loss rates between $10^{-7}$ and $10^{-1}\,\Msunyr$
(following $\rho=\Mdot/4 \pi v_{\rm wind} r^2$, $v_{\rm wind} =100\,{\rm km\,s^{-1}}$) assuming an $r^{-2}$ wind density profile.
For an explanation of the lower limit on the mass-loss rate as obtained by measurement of the H$\alpha$ luminosity, see Methods \S\ref{sec:flux}.
The early-time spectra modeling region is based on the CMFGEN model results (main text and Methods \S\ref{sec:specmodel}).
Also indicated is the overall phase space of Type IIn SNe, based on the compilation appearing in ref.~23, 
where a span of $\Mdot$ between $10^{-4}$ and $10^{-1}\,\Msunyr$ is found to exist based on analyses of spectra
from around 10\,d and onward ($>100$\,d; the corresponding radii assume $v_{\rm shock}=10^4\ {\rm km\,s^{-1}}$).
The top abscissa marks the timespan from explosion for the corresponding radii 
(i.e., when the shock reaches the corresponding radii, applying the above $v_{\rm shock}$).
This schematic figure shows that a CSM with $\Mdot \gtorder 10^{-3}\,\Msunyr$ cannot be part of an extended ($r^{-2}$) wind structure. 
The lack of persistent emission lines for an extended period of time 
(as seen for SNe~IIn), as well as the constraints obtained from the radio nondetections, 
indicate that the dense nearby CSM, with $\Mdot$ around $10^{-3}\,\Msunyr$, must be confined.
Overplotted is a cartoonish visualisation of the proposed CSM structure (the shaded areas convey the span of possible density profiles): 
a nearby, possibly detached, dense shell extending to a radius $\ltorder 10^{15}\,{\rm cm}$,
and a possible underlying extended wind (with $\Mdot \ltorder 10^{-6}\,\Msunyr$) that is not ruled out.
A possible smoother transition between the two components is denoted by the region underlying the question mark.
\label{fig-CSM_config}
}
\end{figure}

We conclude that our study has established that we have observed a regular SN~II that resulted from an explosion of a massive star, 
most likely an RSG, surrounded by dense CSM confined to lie within a few $10^{15}$\,cm from the star (see Fig.~\ref{fig-CSM_config}).
Though several scenarios try to explain the existence of a surrounding confined CSM,
our preferred interpretation is that this material was ejected by the star during the few hundred days prior to its explosion.
As shown in Fig.~\ref{fig-CSM_config}, our observations clearly do not rule out the possible existence of a weaker extended wind, 
expelled by the RSG over a long duration.
Theoretical works (e.g., ref.~9) 
suggest the possible existence of enhanced mass loss, in the form of pulsationally driven superwinds, during the late stages of massive-star evolution.
Therefore, it is possible that the RSG wind profile may be steeper than the assumed $r^{-2}$ profile closer to the star,
creating a smoother transition between the extended wind and the inner dense CSM, as depicted in Fig.~\ref{fig-CSM_config}.
We note the remarkable resemblance of this proposed CSM configuration to the density profile calculated by ref.~29 
(Fig. 1 in that paper) for the inner CSM component of SN 1998S (a type IIn-like SN, which showed more persistent signatures of CSM interaction).
%
Our study suggests that episodic enhanced mass loss by massive stars just prior to their terminal explosion, 
as proposed by several theoretical studies, also occurs among the progenitors of common types of SNe.
This may further strengthen the possibility of a causal connection between the precursor eruptive mass loss
and the ultimate collapse of the core shortly thereafter\cite{2013Natur.494...65O}.

The internal structure of a SN progenitor a short time prior to its collapse is among the major
uncertainties regarding SN explosion modeling, thus strongly motivating further exploration of newborn SNe.
While our study here focuses on a single event with exceptionally early follow-up observations,
available information --- such as the 2009--2014 PTF sample described in ref.~10 
(which includes the Type II SN PTF11iqb also described by ref.~31) 
as well as additional flash-spectroscopy events from the last several years ---
indicates that many core-collapse SNe present combinations of low-expansion-velocity emission lines, 
originating from various elements and ionisation levels, when observed sufficiently soon after explosion.
Future {\em flash-spectroscopy} observations of a larger sample of events would allow us to determine exactly how ubiquitous this phenomenon is, placing stronger constraints on the final stages of massive-star evolution.

%




 \clearpage
 
 \begin{addendum}

 \item[Correspondence] Correspondence and requests for materials 
 should be addressed to Ofer Yaron~(email: ofer.yaron@weizmann.ac.il).


\item
We are grateful to the staffs at the various observatories where data were obtained, 
as well as to N. E. Groeneboom, K. I. Clubb, M. L. Graham, D. Sand, A. A. Djupvik, I. Shivvers, 
J. C. Mauerhan, and A. Waszczak for assistance with observations.
We thank Eli Waxman for valuable discussions. 
AG-Y's group is supported by the EU/FP7 via an ERC grant, the Quantum Universe I-Core program by the Israeli Committee 
for planning and budgeting and the ISF; by Minerva and ISF grants; by the Weizmann-UK ``making connections" program; 
and by Kimmel, ARCHES and Yes awards.
DAP acknowledges support from Hubble Fellowship grant HST-HF-51296.01-A awarded by the Space Telescope Science Institute, and from a Marie Curie Individual Fellowship as part of the Horizon 2020 European Union (EU) Framework Programme for Research and Innovation (H2020-MSCA-IF-2014-660113).
JHG acknowledges support from an AMBIZIONE grant of the Swiss NSF.
EOO is supported by the Arye Dissentshik career development chair, Israel Science Foundation, Minerva, Weizmann-UK, 
and the I-Core program.
MMK acknowledges support from the National Science Foundation for the GROWTH project funded under Grant No 1545949.
AVF's research is supported by the Christopher R. Redlich Fund, the TABASGO Foundation, and US NSF grant AST-1211916.
Support for IA was provided by NASA through the Einstein Fellowship Program, grant PF6-170148.
 LANL participation in iPTF is supported by the US Department of Energy as part of the Laboratory Directed Research and Development program.
Supernova research at the Oskar Klein Centre is supported by the Swedish Research Council and by the Knut and Alice Wallenberg Foundation.
KM acknowledges support from a Marie Curie Intra-European Fellowship, within the 7th European Community Framework Programme (FP7).
 Some data were obtained with the Nordic Optical Telescope, which is operated by the Nordic Optical Telescope Scientific Association at the
Observatorio del Roque de los Muchachos, La Palma, Spain.
We thank the RATIR project team and the staff of the Observatorio Astronomico Nacional on Sierra San Pedro Martir. 
and software support from Teledyne Scientific and Imaging. 
RATIR, the automation of the Harold L. Johnson Telescope of the Observatorio Astron«omico Nacional on Sierra San Pedro Martir, and the operation of both are funded through National Aeronautics and Space Administration (NASA) grants NNX09AH71G, NNX09AT02G, NNX10AI27G, and NNX12AE66G, CONACyT (INFR-2009-01-122785, CB-2008-101958), UNAM PAPIIT (IN113810 and IG100414), and UCMEXUS-CONACyT.
Some of the data presented herein were obtained at the W. M. Keck Observatory, which is operated as a scientific partnership among the California Institute of Technology, the University of California, and NASA.
The Observatory was made possible by the generous financial support of the W. M. Keck Foundation.
Research at Lick Observatory is partially supported by a generous gift from Google.
A portion of this work was carried out at the Jet Propulsion Laboratory under a Research and Technology Development Grant, 
under contract with NASA.

\end{addendum}



\clearpage

\newpage

\newpage

\newpage

\newpage

\clearpage

\begin{methods}

In this section we describe the data, methods, and theoretical calculations used in the main paper.
This work presents our photometric and spectroscopic observations of \dqy\ (\fs) during the first two months from discovery.
PTF $r$-band observations were obtained by the PTF survey camera mounted on the Palomar 48-inch Schmidt telescope\cite{2009PASP..121.1395L, 2009PASP..121.1334R}.
Photometry is measured using our custom pipeline performing point-spread-function (PSF) photometry on PTF images after removing a reference image constructed from pre-explosion data using image subtraction.  
Using the multi-band UVOIR photometric measurements, we constructed fits to the spectral energy distribution (SED) and derived
the evolution of the blackbody (BB) temperatures and radii between a few hours and $2$ months after explosion.
To derive estimates of the progenitor's mass-loss rate, the effective temperature, and radius of the emitting (CSM) region,
we utilised the radiative-transfer code CMFGEN\cite{1998ApJ...496..407H} to perform detailed spectroscopic modeling of the early-time spectra.
We undertook radio observations of \dqy\ with the Jansky Very Large Array (VLA) at $\sim70$ and 100\,d after explosion.
These observations, at central frequencies of 6.1\,GHz and 22\,GHz, resulted in null detections, which can be translated into limits on the CSM density.
Finally, we derived estimates for the radius and energy per unit mass of the exploding star using the formalism of RW11, 
and discussed calculations of the SBO using custom improved models.

\section{Discovery}
\label{sec:disc}

The intermediate Palomar Transient Factory 
(iPTF\cite{2013ATel.4807....1K}, which began operating in 2013 as a continuation of PTF\cite{2009PASP..121.1334R, 2009PASP..121.1395L}), 
utilises the 48-inch Samuel Oschin telescope (P48) at Palomar Observatory, California, USA to monitor the transient sky.
Each P48 frame, consisting of 11 CCD chips, has a wide field of view of $7.2$ square degrees\cite{2008SPIE.7014E..4YR}.
Fields are monitored with a short cadence (1--2\,d), and 
at least two images are taken per field per night, separated by $\gtorder30$ min, to search for transient events.
The images are processed\cite{2014PASP..126..674L, 2012PASP..124...62O, 2012PASP..124..854O}, 
a reference image is subtracted, and candidate transients are selected with the aid of machine-learning algorithms\cite{2013MNRAS.435.1047B, 2013AAS...22143105W, 2015AAS...22543402R} for distinguishing a real astrophysical source from bogus artifacts.
A shorter list of candidates is then vetted by human ``scanners"\cite{2011ApJ...736..159G} who save the candidates of interest and assign prioritised follow-up observations.

\dqy\ was first detected on 2013 Oct. 6.245. Following the second detection confirming the discovery $\sim50$ min later (Oct. 6.279), it was automatically saved by a robotic process (internally referred to as ``the treasurer") on Oct. 6.365. 
The last nondetection at its location, at a limiting magnitude of $m_r=21$ in the PTF $r$ band, is from 22\,hr before the first detection.
A follow-up campaign was promptly initiated by the astronomer on duty, including a very early sequence of spectra at the Keck-I telescope, beginning a mere half hour after the candidate was saved. 

The P48 discovery image and a Sloan Digital Sky Survey (SDSS) colour image of the location and host galaxy are shown in Supplementary Fig.~\ref{SIfig-discovery}.

\section{Photometry}
\label{sec:phot}

Shortly after the discovery with the P48, we initiated an extensive follow-up campaign; the multiband light curves are presented in Fig.~\ref{fig-photometry}.
Less than 3\,hr after the first P48 detection, we began obtaining photometry with the 
Palomar 60-inch telescope (P60; ref.~44) 
in the $g$, $r$, and $i$ filters.
Beginning 17\,hr after the first detection, {\it Swift}-UVOT photometry was obtained with filters $UVW2$, $UVM2$, and $UVW1$, 
accompanied by UVOT $U$, $B$, and $V$ photometric measurements from day $3.6$ onward.
An LCOGT multiband photometry campaign was initiated around 1\,d after discovery, while
RATIR observations (in filters $i$, $Z$, $Y$, $J$, and $H$) began just prior to day 2 after discovery.

Observations in the $r$ band were obtained by the iPTF survey camera mounted on the P48 telescope\cite{2009PASP..121.1395L, 2009PASP..121.1334R}.
Magnitudes were measured using our custom pipeline performing point-spread-function (PSF) photometry on iPTF images after removing a reference image constructed from pre-explosion data using image subtraction.  
{\it Swift} Ultraviolet absolute AB magnitudes were measured using standard pipeline reduction and are corrected for host-galaxy contamination using late-time {\it Swift} images.
All measurements were corrected for Galactic (Milky Way) extinction using the ref.~45 
reddening law (assuming $R_V=3.08$).

To remove contamination from the underlying host galaxy, SDSS frames were subtracted from the P60 images using a similar technique to that applied to P48 data.  
The resulting images of the SN were then photometrically calibrated with respect to several SDSS point sources within the P60 field of view.

LCOGT magnitudes were obtained using PSF fitting after background removal using a low-order polynomial fit, without template subtraction.
The LCOGT $U$, $B$, $V$, $R$, and $I$ magnitudes were converted from the Vega to 
AB systems using the conversion values as specified in Table 1 of ref.~46. 
The RATIR $J$ and $H$ magnitudes were converted from Vega to AB according to the conversion values given in Table 7 of ref.~47.

The {\it Swift}-XRT (ref.~48) 
observations, conducted between days 1 and 25 after explosion (with a roughly constant cadence of 1--2\,d), 
were reduced using the tools of ref.~49, 
applying a $9''$ aperture radius centred on the SN position and correcting for 50\% flux losses.
We examined several binning schemes relative to the estimated explosion time, ${\rm JD_0} =$ 2,456,571.62.
Binning the measurements into three segments, we find
two marginal ($<3\sigma$) detections and an upper limit as follows (in counts per kilosecond, specifying 1$\sigma$ and 2$\sigma$ errors):

\noindent\newline$<t> = 3.4$\,d [0--5],         $0.42_{-0.23,-0.33}^{+0.40,+0.82}$ ct/ks,
\noindent\newline$<t> = 11.0$\,d [5--15],     $<0.67$ (2$\sigma$) ct/ks, and
\noindent\newline$<t> = 20.3$\,d [15--25],   $0.38_{-0.21,-0.30}^{+0.37,+0.75}$ ct/ks.

\noindent\newline Assuming a power-law spectrum with a photon index of 2, and 
Galactic extinction $N_{\rm H}=3.9\times 10^{20}\, {\rm cm^{-2}}$ toward the SN location, 
the conversion of count rates to flux follows $1\,{\rm ct/s}\ \approx 4\times 10^{-11}\,{\rm erg\,cm^{-2}\,s^{-1}}$, giving X-ray luminosities of 
$L_{\rm X}=6.56\times 10^{40}$, $<1.05\times 10^{41}$, and $5.94\times 10^{40}$\,erg\,s$^{-1}$, respectively (applying a distance of 50.95 Mpc).
However, these are all marginal ($<3\sigma$) detections, and the flux could originate from noise fluctuations or the galaxy rather than the SN.
Future deep observations may provide a better background estimate.


A coadd of all data points provides a one-sided 95\% confidence limit of $<3\times10^{-4}$ ct/s, which corresponds to an upper limit on the X-ray luminosity of $L_X \ltorder 4.7\times 10^{40}\,{\rm erg\,s^{-1}}$.

\section{Spectroscopy}
\label{sec:spec}

All spectra were reduced using standard pipelines.
Our earliest (6--10\,hr) and latest (57\,d) spectra were obtained using the Low Resolution Imaging Spectrometer 
(LRIS\cite{1995PASP..107..375O}; Keck observations PI D.~Perley) mounted on the Keck-I 10\,m telescope (using the 600/4000 grism and 400/8500 grating).
A careful procedure was applied for subtracting the flux of the host from the four early-time Keck spectra,
involving the subtraction of an offset H~II region spectrum in each frame.
As a result of this, the narrow-component emission lines in the early Keck spectra appearing in Fig.~\ref{fig-earlyspec}
originate from the flash-excited circumstellar emission and are not contaminated by the underlying host H~II emission. 
A higher-resolution spectrum was also obtained with Keck-I/LRIS (using the 1200/7500 grating) 
on the first night, at $\sim 10.3$\,hr after the estimated explosion time. 
A section of this spectrum centred on the H$\alpha$ line is shown in Supplementary Fig.~\ref{SIfig-halpha_hires}, 
together with a combination of Gaussian and Lorentzian fits to the narrow and wide components, respectively.
The instrumental full width at half-maximum intensity (FWHM) is $\sim2.35$\,\AA\ (from analysis of night-sky lines and of the background nebular H$\alpha$ emission, and consistent with the quoted resolution of the grating), so the narrow component is likely unresolved down to $\sim100\ {\rm km\,s^{-1}}$ 
(thus placing an upper limit on the velocity dispersion of the emitting material).

Additional spectra were obtained using the Double Beam SPectrograph (DBSP\cite{1982PASP...94..586O})  mounted on the 5.1\,m Palomar Hale (P200) telescope, ALFOSC mounted on the 2.56\,m Nordic Optical Telescope (NOT), FLOYDS mounted on the 2\,m Faulkes Telescope South (FTS) (day 3), DEep Imaging Multi-Object Spectrograph (DEIMOS\cite{2003SPIE.4841.1657F}) mounted on the Keck-II 10\,m telescope, DIS mounted on the 3.5\,m Apache Point Observatory (APO) telescope, ISIS mounted on the 4.2\,m William Herschel Telescope (WHT), and Kast\cite{miller1994kast} mounted on the 3\,m Lick Shane telescope.

Fig.~\ref{fig-latespec} displays the later spectra, extending from day 8 through $\sim2$ months after explosion, revealing  developed P-Cygni profiles typical of spectroscopically normal SNe~II.
Evolution of the expansion velocity of the SN ejecta from the absorption P-Cygni features of several species is shown in Supplementary Fig.~\ref{SIfig-velevol}.

The log of spectroscopic observations is presented in Table~\ref{SItab-spec}.
All spectra and their accompanying meta-data are publicly available via WISeREP\cite{2012PASP..124..668Y} (http://wiserep.weizmann.ac.il).

\section{Bolometric Luminosity and Blackbody Fits}
\label{sec:bbfits}

Using the multiband UVOIR photometric measurements interpolated to a common grid, 
we constructed fits to the SED and derived
the evolution of the blackbody (BB) temperatures and radii 
between $\sim3$\,hr (the first P48 detection) and 2 months after explosion.
It should be emphasised that the treatment as a BB is an approximation; 
during the first several hours after explosion the SEDs seem to roughly follow a modified BB spectrum 
below the peak, $L_\nu\propto\nu$ or $L_\lambda\propto\lambda^{-3}$.

For each point in the temporal grid, the best-fit BB temperature was obtained as follows.
For each trial temperature, within a scanned range of temperatures, we calculate the synthetic photometry
of the BB spectrum, in the specific bands for which we have observations.
Then we can calculate the root-mean square (RMS) difference between the observed magnitudes
and the calculated (synthetic) magnitudes and look for the best-fit BB temperature.
For each derived best BB temperature, the best-fit angular radius can be obtained,
and this is turned into the physical radius (cm) by applying the known distance to the SN.

These values are plotted in Supplementary Fig.~\ref{SIfig-BBTempRad}, 
with the temperature estimate obtained from the modeling of the earliest spectrum (described in Methods \S\ref{sec:specmodel}) overplotted.
In order to investigate the earliest emission (beginning with the first P48 $r$-band detection), 
we carefully extrapolate the P60 $g$ and $i$ and the UVOT $UVM2$ light curves 
(for which we have observations beginning $\sim 3$ and 17\,hr after the first detection, respectively) 
back to the first P48 detection.
The BB temperature estimate for this early point is in excess of 100\,kK and should be taken with due caution; 
we estimate its uncertainty range at 60\,kK.

In Supplementary Fig.~\ref{SIfig-BolLum}, we display estimates of the bolometric flux as obtained by three methods.
The most conservative lower limit on the bolometric luminosity is based on our sequence of optical spectra, 
by integration of the total flux under each spectrum (and within the wavelength range covered by the spectrum).
Next, we estimated the bolometric light curve based on the multiband photometric measurements, 
using the same interpolated common grid mentioned above.
Experimenting with different fits and interpolation schemes of the SED (see figure caption), 
based on the observed magnitudes at the different bands, produces the span shown by the grey shaded area.
Because our multiband (UVOIR) photometric measurements cover a wider wavelength range 
during most periods of these first 2 months after explosion, these bolometric flux estimates set a tighter lower limit to the actual bolometric light curve.
Finally, we plot a bolometric luminosity light curve based on the derived BB temperatures and radii estimates, 
as described above, by applying $L_{\rm bol}=4\pi R_{\rm BB}^2\sigma T_{\rm BB}^4$.
According to these BB estimates, the bolometric luminosity at $\sim 3$\,hr after explosion surpasses $10^{44}\ {\rm erg\,s^{-1}}$. 
It can also be seen in the figure that these values are in good agreement with the previous $L_{\rm bol}$ estimates, based on the multiband photometry, especially from around day 4 onward.

\section{Line Fluxes}
\label{sec:flux}

In addition to the other estimates of the mass-loss rate and the radius of the emitting (CSM) region, described in the main text,
we can also obtain an order-of-magnitude estimate for the mass loss based on the measured Balmer H$\alpha$ luminosity, 
following the expressions given by ref.~54. 

A basic underlying assumption is that the CSM around the progenitor has a spherical wind density profile of the form $\rho = Kr^{-2}$,
where $r$ is the distance from the progenitor and $K \equiv \Mdot / (4\pi v_{\rm wind})$ is the mass-loading parameter ($\Mdot$ being the mass-loss 
rate and $v_{\rm wind}$ the wind expansion velocity).

Based on the maximal H$\alpha$ line flux (Supplementary Fig.~\ref{SIfig-lineflux}), $f_{{\rm H}\alpha}=6.25\times 10^{-15} {\rm \,erg\,s^{-1}\,cm^{-2}}$
as measured from the Oct. 7 (day 1.4 from explosion) FTS spectrum, the H$\alpha$ luminosity is\\
$L_{{\rm H}\alpha}=f_{{\rm H}\alpha}4\pi d^2=1.94\times 10^{39}{\rm \,erg\,s^{-1}}$, where $d=50.95$ Mpc is the luminosity distance to the SN.

According to Eq.~6 in ref.~54, 
a relation can be obtained between the mass-loading parameter $K$, the H$\alpha$ luminosity, and the radius:
$L_{{\rm H}\alpha}\ltorder A K^2 / r$, 
where $A=4\pi h\nu \alpha_{\rm H}^{\rm eff}/(<\mu_p>m_{p}^{2})$;  
$<\mu_p>=0.6$ is the mean molecular weight, and
$\alpha_{\rm H}^{\rm eff}\approx 8.7\times 10^{-14}(T_{\rm eff}/(10^4\,{\rm K})^{0.89}$ (ref.~55). 
For optical depth $\tau \equiv \kappa \rho r \approx 1$ ($\kappa$ taken to be $0.34$), $K\approx r/\kappa$ (for the assumed wind density profile $\rho = Kr^{-2}$),
and we obtain\\
$r \gtorder \kappa ^2 L_{H\alpha} / A \approx 1.1\times 10^{14}$\,cm, and\\ 
$\rho = 1/(\kappa r) \approx 2.6\times 10^{-14} {\rm \,g\,cm^{-3}}$, 
corresponding to a particle density of\\
$n=\rho / (\mu_p m_p)\approx 2.6\times 10^{10} {\rm \,cm^{-3}}$.

The derived mass-loading parameter is $K=\rho r^2\gtorder 3.3\times 10^{14}\ {\rm g\,cm^{-1}}$, 
which leads to a lower-limit estimate of the mass-loss rate of
$\Mdot = 4\pi K v_{\rm wind} \gtorder 7\times 10^{-4} (\frac{v_{\rm wind}}{100\ {\rm km\,s^{-1}}}) {\rm \,\Msun\,yr^{-1}}$.


Additional lower limits on the mass-loss rate can be placed by analysis of the electron-scattering wings seen in the emission lines
during the first several days. 
Following ref.~56 
and references therein, the optical depth to electron scattering of the wind can be expressed as\\
$\tau_{\rm wind} = 1.16\Big(\frac{\Mdot}{10^{-3}\, \Msunyr}\Big)\Big(\frac{v_{\rm wind}}{100\ {\rm km\,s^{-1}}}\Big)^{-1}\Big(\frac{v_{\rm shock}}{2 \times 10^4\ {\rm km\,s^{-1}}}\Big)^{-1}\ t_{\rm days}^{-1}\, {\rm cm}^{-2}$.

The fact that the H$\alpha$ line is dominated by electron scattering over the first few days,
which requires $\tau_{\rm elec. scatt.} \gtorder 2$--3,
therefore gives an independent mass-loss rate estimate of $\Mdot \gtorder 2 \times 10^{-3}\ \Msunyr$ (for the assumed wind velocity).

Supplementary Fig.~\ref{SIfig-lineflux} presents the manner by which the major observed emission lines of the flash-ionised spectra disappear within the first several days after explosion.
The  H$\alpha$, H$\beta$, and He~II lines first increase in flux, reaching a maximum around day 1 from explosion. 
The He~II line disappears between days 2 and 3, whereas the highly ionised oxygen lines all disappear within the first day.

\section{Emission-Line Spectra Models}
\label{sec:specmodel}

We performed detailed spectroscopic modeling of the early-time spectra of \dqy\ using the radiative-transfer code CMFGEN\cite{1998ApJ...496..407H} and the same model assumptions as described by ref.~20. 
The free parameters are essentially the boundary of the inner radius ($R_{\rm in}$) 
and the bolometric luminosity at this inner boundary ($L_{\rm SN}$).
$L_{\rm SN}$ and  $R_{\rm in}$, which determine the effective temperature ($T_{\rm in}$) at the inner boundary, 
govern the ionisation structure of the illuminated progenitor wind (the post-SN spectrum) and also the absolute flux level.
Having a flux-calibrated spectrum allows us to determine both $L_{\rm SN}$ and $R_{\rm in}$ (i.e., they are not assumed {\it a priori}), since models with the same $R_{\rm in}$ but different $L_{\rm SN}$ will have different absolute fluxes and ionisation structures (i.e., different spectral line ratios).
The other free parameters are the progenitor $\Mdot$, $v_{\rm wind}$ (assumed constant, since no hydrodynamical modeling is performed), 
and the chemical composition. 
The resulting values described below were all obtained for models applying $v_{\rm wind}=100\,{\rm km\,s^{-1}}$.
The best-fitting models (shown in the figures) were obtained for a He-enriched surface composition 
($Y=0.49$, $X=0.49$, for the helium and hydrogen mass fractions, respectively); 
the rest of the abundances are consistent with solar.

Fig.~\ref{fig-Groh_spec} displays the comparison of the obtained model spectra to the first Keck spectrum at $\sim6$\,hr after the explosion.
The three model spectra plotted in the figure are the results for the following parameter combinations:\\
Red: $R_{\rm in}=1.40\times10^{14}\,{\rm cm}$,  $L_{\rm SN}=2.00\times10^{10}\,{\rm \Lsun}$ ($T_{\rm in}=48.4\,{\rm kK}$), $\Mdot=2\times10^{-3}\,{\rm \Msun\,yr^{-1}}$;\\
Blue: $R_{\rm in}=1.33\times10^{14}\,{\rm cm}$,  $L_{\rm SN}=2.75\times10^{10}\,{\rm \Lsun}$ ($T_{\rm in}=53.5\,{\rm kK}$), $\Mdot=3\times10^{-3}\,{\rm \Msun\,yr^{-1}}$;\\
Orange: $R_{\rm in}=1.30\times10^{14}\,{\rm cm}$,  $L_{\rm SN}=3.50\times10^{10}\,{\rm \Lsun}$ ($T_{\rm in}=58.5\,{\rm kK}$), $\Mdot=4\times10^{-3}\,{\rm \Msun\,yr^{-1}}$.

As is evident from the plots, these three models bracket the observed spectra quite well and are the best obtained fits. 
One of the main issues is to reproduce the O~VI $\lambda\lambda3811$, 3834 and the O~V $\lambda5597$ features simultaneously. This is crucial because it is the only temperature indicator we have in the optical region (we need lines of the same species but different ionisation stages). The O~V emission needs $T_{\rm in}\ltorder49$\,kK to be consistent with the observations, but little O~VI is present then. O~VI requires $T_{\rm in}\gtorder53$\,kK, but then O~V is too weak. These values give a range of temperatures that could be consistent with the data given our model assumptions --- i.e.,  $T_{\rm in}\approx48$--58\,kK.
A multitemperature configuration may be required to further improve the fits.

$\Mdot$ is mainly determined by the strength of the ${\rm H\alpha}$ and He II $\lambda4686$ lines. It varies according to the choice of $T_{\rm in}$, 
so values around (2--4) $\times10^{-3}\ {\rm \Msun\,yr^{-1}}$ are consistent with the observations. 
These assume $v_{\rm wind}=100\ {\rm km\,s^{-1}}$ and are thus upper limits;
$\Mdot$ can easily be scaled  down for lower $v_{\rm wind}$ by multiplying the above $\Mdot$ range
by ($v_{\rm wind}/100\ {\rm km\,s^{-1}}$), similar to the scaling expression specified in Eq. 1 of ref.~20. 

As mentioned, because the wind velocity of $100\,{\rm km\,s^{-1}}$ is in practice an upper limit,
additional models were tested applying a low wind velocity of $15\,{\rm km\,s^{-1}}$, more typical for standard RSG winds.
The resulting estimates of the mass-loss rate are an exact match with the above scaling relation, 
the $\Mdot$ range drops down to (3--6) $\times10^{-4}\ {\rm \Msun\,yr^{-1}}$,
with $R_{\rm in}=(1.3-1.4) \times 10^{14}\,{\rm cm}$ and $L_{\rm SN}=(2.2-3.7) \times 10^{10}\,{\rm \Lsun}$ in order to match the observed flux.
We do argue, however, that a significantly elevated mass-loss rate over a short time (as we argue is the case here) 
can be caused by different physical mechanisms than those driving a ``normal" wind; 
thus, assuming higher velocities, up to $100\,{\rm km\,s^{-1}}$, is reasonable.

The final models (shown in Fig.~\ref{fig-Groh_spec}) were obtained for an enhanced He abundance of $Y=0.49$;
however, such enhanced surface He abundances ($\gtorder0.40$) are still consistent with 
solar initial abundances (e.g., refs~57,58). 
Fe is assumed to be solar because there are no Fe lines. 

The enhancement of helium seems to be a pretty robust outcome from our modeling, 
but this should also involve an enhancement in the surface nitrogen abundance.
Although the possible nitrogen lines identified in the early spectra of \dqy\ are quite weak compared to those of
SN\,2013cu\cite{2014Natur.509..471G} and SN\,1998S 
(for which ref.~16 
recently performed a thorough examination of an early-time (few days) Keck-I HIRES spectrum), 
the fact that the strength of the N~V line is extremely sensitive to the temperature,
as well as the location on top of the strong, asymmetric electron-scattering wings of the He~II $\lambda4686$ line,
means that the nitrogen abundance can be consistent with the required He enhancement.  
With the highly resolved SN\,1998S spectrum\cite{2015ApJ...806..213S} serving as a reference, 
we verify that the ``shoulder" that develops blueward of the He~II $\lambda4686$ in the \dqy\ spectra between $\sim10$ and 21\,hr 
does not emanate from N~III $\lambda\lambda$4634, 4641,
but rather from N~V $\lambda\lambda$4604, 4620.

The models do not allow He~II $\lambda4686$, ${\rm H}\alpha$, and O~VI simultaneously at $T_{\rm in}\gtorder90$\,kK with solar abundance. 
In high effective temperature models ($T\gtorder100$\,kK), the oxygen abundance has to be increased by a factor of 10 to get the O~VI line, 
but then our models are not able to reproduce the O~V and O~IV lines.

We note that whereas the identification of the O~VI line is secure and the shape of the double peak matches the models well, 
the identification of O~V $\lambda5597$ is somewhat less certain.
This line is in a region of the spectrum that is affected by the dichroic and the overall shape of the feature is contaminated by an instrumental artifact, 
thus not matching the exact line shapes as seen for the other emission lines and as obtained by the models. 
However, O~V is a key line, and since it shows up in other models in different parameter spaces (for example, WO Wolf-Rayet stars, for which the emission of ionized oxygen dominates), we would be surprised if the observed feature does not contain an O~V line that is of real origin.

The feature around 5800\,\AA\ (redward of the O~V line) that is present in the models is C~IV, 
and the feature around 3400\,\AA\ is O~IV.
A possible explanation for the complete nonexistence of the C~IV line in the observed early-time spectra 
could be that the carbon abundance is slightly lower than solar.

With relation to the line-forming regions (Supplementary Fig.~\ref{SIfig-Groh_line_form}), 
both O~V and O~VI originate partially in areas of high Thomson opacity ($\tau_{\rm Thomson} \approx 1$--2),
leading to electron-scattering wings that are relatively stronger compared to, say, ${\rm H}\alpha$. 


Finally, we note that following the comparison of the early-time spectrum of SN\,2013cu 
to Wolf-Rayet (WR) models as done by ref.~2, 
here we also examined the PoWR grid of WR model spectra (http://www.astro.physik.uni-potsdam.de/~wrh/PoWR/powrgrid1.php) 
with respect to the early flash-ionised spectra of \dqy. 
While such a comparison is useful to provide a handle on line identification and possible temperature regimes,
our conclusion is that a direct comparison (to the existing WR ``families" of models) is not applicable for this particular case.
The WR high-temperature models that contain hydrogen are also nitrogen rich (e.g., WNL-H50),
and the carbon/oxygen-rich models (WC/O) all have very prominent C~IV lines that are not observed in our spectra.

\section{Radio Analysis}
\label{sec:radio}

The interaction between SN ejecta and surrounding material can produce synchrotron
emission; thus, radio observations can provide powerful diagnostics of
the CSM\cite{1982ApJ...259..302C, 1998ApJ...499..810C, 2006ApJ...651..381C, 2002ARA&A..40..387W}.
The radio emission can be used to constrain physical properties 
such as the CSM density and the CSM shockwave radius and velocity.

We observed \dqy\ with the Jansky Very Large Array (VLA) on
2013 Dec. 17 (PI A.~Horesh). The observation was undertaken in both the C and
K bands (at central frequencies of 6.1\,GHz and 22\,GHz,
respectively). Data reduction was performed using the AIPS\footnote{http://www.aips.nrao.edu}
software\cite{2003ASSL..285..109G} with 3C~48 as a flux calibrator and J2330+11 as
a phase calibrator. The observations resulted in a null detection with
RMS values of 17\,$\mu$Jy and 14\,$\mu$Jy in the C and K bands, respectively. 
A second observation took place on 2014 Jan. 18 (with the same configuration and setup),
resulting again in a null detection with an RMS of 12\,$\mu$Jy and 10\,$\mu$Jy in the C and K bands, respectively. 

The above null detections can now be translated into limits on the CSM
density. Adopting the synchrotron self-absorption model of ref.~60, 
and assuming equipartition, we can calculate the expected radio
emission in both the C and K bands. We assume a continuous CSM which is
a result of a stellar wind with constant mass-loading parameters, 
$A \equiv \dot M/(4\pi v_{\rm wind})$.
In the context of a model where the CSM is created by a constant wind, our
radio observations rule out a large subset of mass-loading parameters.
 
We find that a mass-loss rate in the range $6\times 10^{-6} \ltorder\dot M \ltorder 10^{-3}\,{\rm M_{\odot}\,yr^{-1}}$  
cannot be sustained out to a distance of $\sim 10^{16}$\,cm that we probe with the radio observation.
The high end of this range ($\sim 10^{-3}\,{\rm M_{\odot}\,yr^{-1}}$) 
is due to the large amount of free-free absorption.
This value is obtained by assuming a low electron temperature of $2\times 10^4$ K, as may be evident from our models. 
However, there is uncertainty in determining the electron temperature at large distances at later times. 
The free-free optical depth, which highly depends on the electron temperature ($\tau_{\rm ff} \propto T_{e}^{-1.5}$), 
may therefore decrease if a higher electron temperature is assumed to exist in the area probed by the radio observations. 
Past studies have shown that the electron temperature at these distances can indeed be as high 
as $10^{5}$--$10^{6}$\,K (e.g., ref.~64). 
For such higher electron temperatures, the excluded mass-loss rate can be as high as $\sim 10^{-2}\,{\rm M_{\odot}\,yr^{-1}}$,
thus providing a tighter constraint to the confinement of the CSM at the derived mass-loss rates.

\section{Shock-Breakout Analysis}
\label{sec:sbo}

We begin by considering the relevant timescales for the early observations.
The little to no evolution of the recombination lines (Fig.~\ref{fig-earlyspec}) suggests a timescale of $> 4$\,hr (see above).
In the proposed SBO scenario, the breakout burst ionises the material above the progenitor's edge, which later produces recombination lines.
A specific timescale for the recombination lines can be attributed to the longest timescale among the duration of the ionisation burst --- 
the recombination time or the light-travel timescales.
Considering that for the deduced densities, the recombination time is of order minutes\cite{2014Natur.509..471G}, and that the ionisation burst timescale is usually shorter than the 
light-travel time in a spherically symmetric explosion, we attribute the recombination lines timescale to the light-travel time of an extended shell.

However, in such a case the continuum flux probably has a different origin than the recombination lines, such as coming from inside the stellar edge where the SBO occurs.
If both the continuum flux and the recombination lines are to originate at the extended shell, either the emission is produced by a stellar SBO and is then ``smeared" by the light-travel time of an extended shell, or the SBO is inside the extended shell itself.
The problem with the first option is that the total energy released in the burst is dictated by the stellar radius, but the burst timescale is determined by the extended shell radius, and the observed flux is lower compared to a stellar SBO.
The observed flux is too high to consider such an option.
The problem with the second option is that for the SBO to occur inside the extended shell, the shell must be optically thick 
(with an optical depth $>30$), but the line-scattering wings imply an optical depth of only 2--3.

We used our multigroup radiation-hydrodynamics code (Szabo, Sapir, \& Waxman, in prep.) to simulate the continuum spectral emission generated by SBOs from simple spherical stellar envelopes and from envelopes surrounded by optically thick steady mass-loss wind. For the latter,  we tested both truncated and untruncated winds and found that for parameters of the order deduced from observations of \dqy, there is persistent X-ray emission starting several hours (and up to one day) after SBO, owing to the interaction between the post-breakout collisionless shock and the CSM within which it propagates. This X-ray emission is not completely reprocessed by the ionised hydrogen of the CSM and we find that a luminosity of $\sim10^{42 }$--$10^{43}$\,erg\,s$^{-1}$ should be detected by the {\it Swift}-XRT, had this been the case. The full spectral analysis implies that this event cannot have been an SBO from within an optically thick ($\tau \gtorder 30$) shell of steady mass-loss wind, regardless of its outward extension from the stellar surface (as long as it is sufficiently optically thick).

For completeness, we applied the same code to SBOs from a simple spherical stellar envelope (i.e., without CSM at all) and followed them to the late spherical expansion phase. The continuum emission spectra from 6--10\,hr after explosion is best fit by a model of a $10{\rm \,\Msun}$ RSG envelope with breakout shock velocity of  $0.1c$ and a stellar radius of $1000{\rm \,\Rsun}$. The calculated models consider only fully ionised hydrogen, so the above quoted values should be considered as upper limits (cf. the RW11 model, which includes ionisation of He and C/O envelopes). It is therefore plausible that the continuum emission originates from the SBO at the stellar surface, while the spectral emission lines come from an optically thin ($\tau \approx$ a few) CSM shell ionised by the SBO flash.
However, a full model, integrating SBO, cooling, and radiative transfer, as well as the possible effects of asphericity,
may be required to fully recover the physics from the available extensive observations.

\subsection{Data Availability Statement}
The photometry and spectra of \dqy\ (\fs) presented in this study are available from WISeREP\cite{2012PASP..124..668Y}: https://wiserep.weizmann.ac.il.

\end{methods}








\clearpage

\begin{SIfigure}
\centering
\includegraphics[width=11cm]{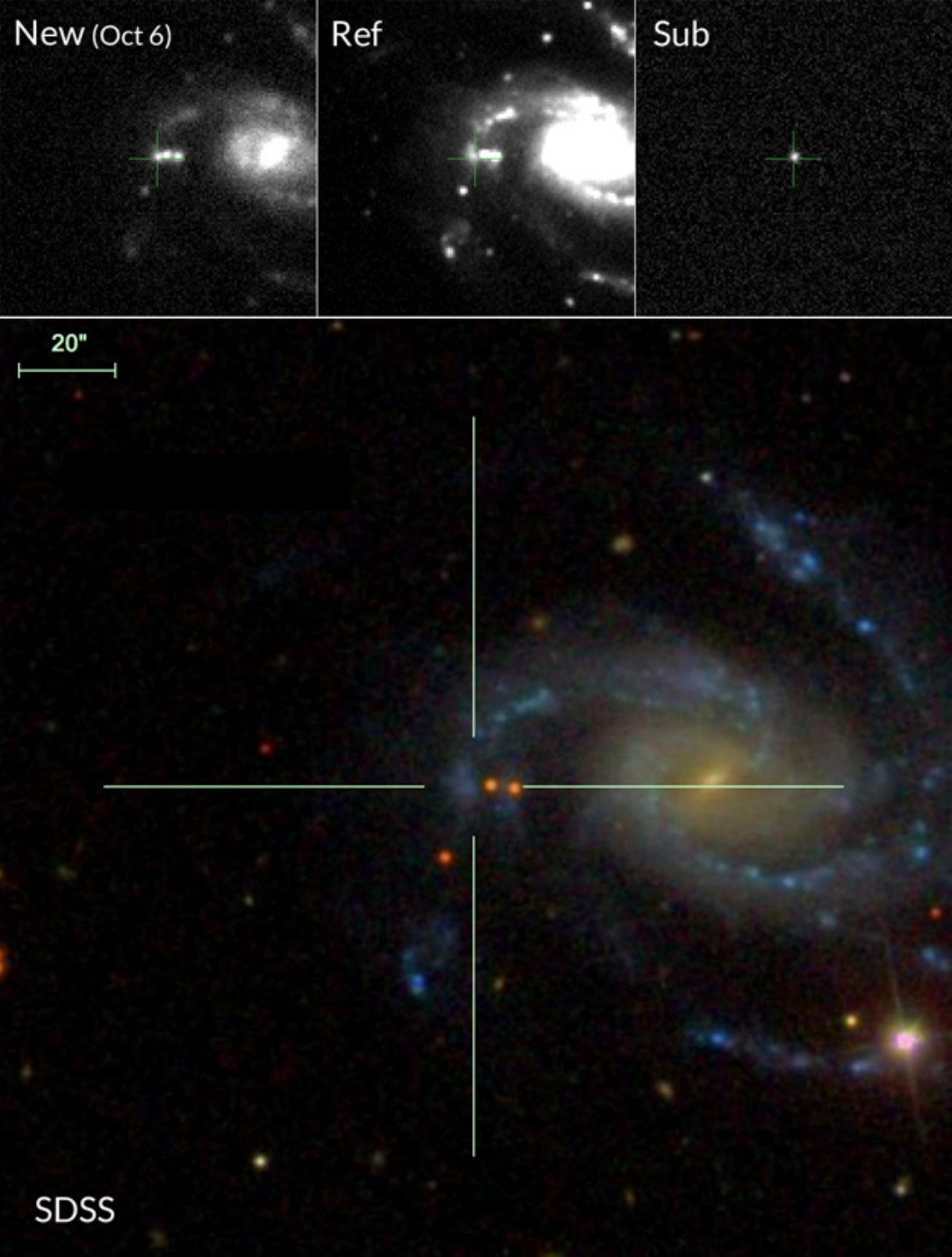}
\caption{Discovery of \dqy\ in the nearby galaxy NGC\,7610 ($d = 50.95$\,Mpc), at $\alpha =$ 23$^h$19$^m$44.70$^s$, $\delta = +10^\circ 11' 04.4''$ (J2000.0).
Top: Palomar 48-inch sequence of the new discovery image from 2013 Oct. 06.24, a reference image (a coadd of pre-explosion images), and the subtraction image.
Bottom: The colour SDSS image.
The SN is located in a blue, star-forming area (the red point sources in the vicinity are foreground stars), 
which is apparently a part of one of the major arms of the spiral host.
\label{SIfig-discovery}}
\end{SIfigure}

\begin{SIfigure}
\centering
\includegraphics[width=17cm]{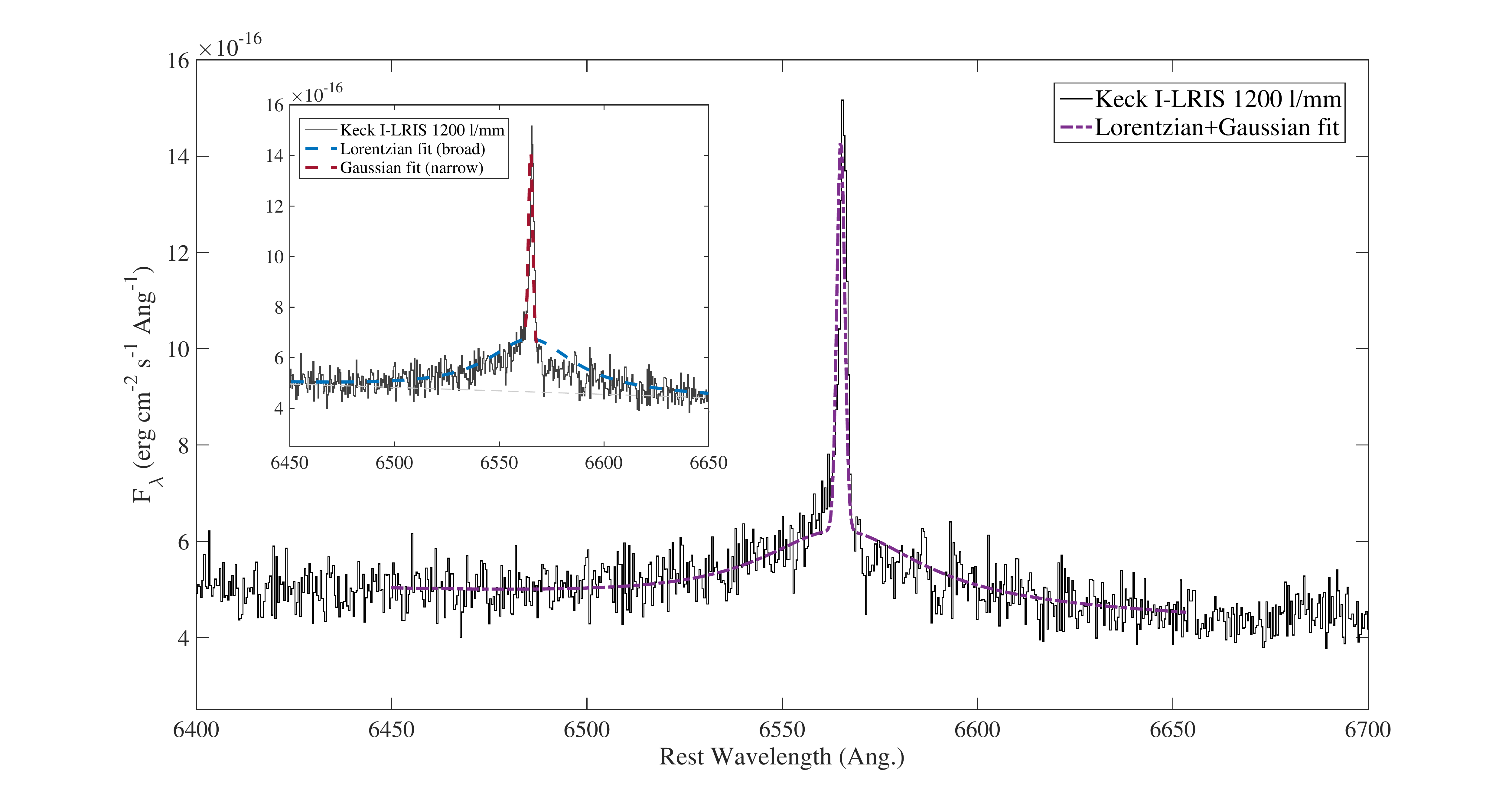}
\caption{Composite Gaussian and Lorentzian fits to the narrow and broad components of the H$\alpha$ line in the Keck-I/LRIS 1200 lines\,mm$^{-1}$ spectrum obtained $\sim10.3$\,hr after explosion. The FWHM of the narrow Gaussian fit is $\sim2.35$\,\AA, corresponding to a wind velocity of $v_{\rm wind}\approx100\ {\rm km\,s^{-1}}$.
The instrumental resolution is around this velocity (LRIS manual, and verified from night-sky lines), so
the line should be regarded as barely resolved, and the velocity estimate should be taken as an approximate close upper limit.
The FWHM of the underlying broad wings, which we relate to electron scattering, is $\sim25$\,\AA. 
The inset displays separate Lorentzian and Gaussian fits to the two components, focusing on the possible asymmetry of the line profile, 
especially a potential lack of flux on the red side.
\label{SIfig-halpha_hires}}
\end{SIfigure}

\begin{SIfigure}
\centering
\includegraphics[width=16cm]{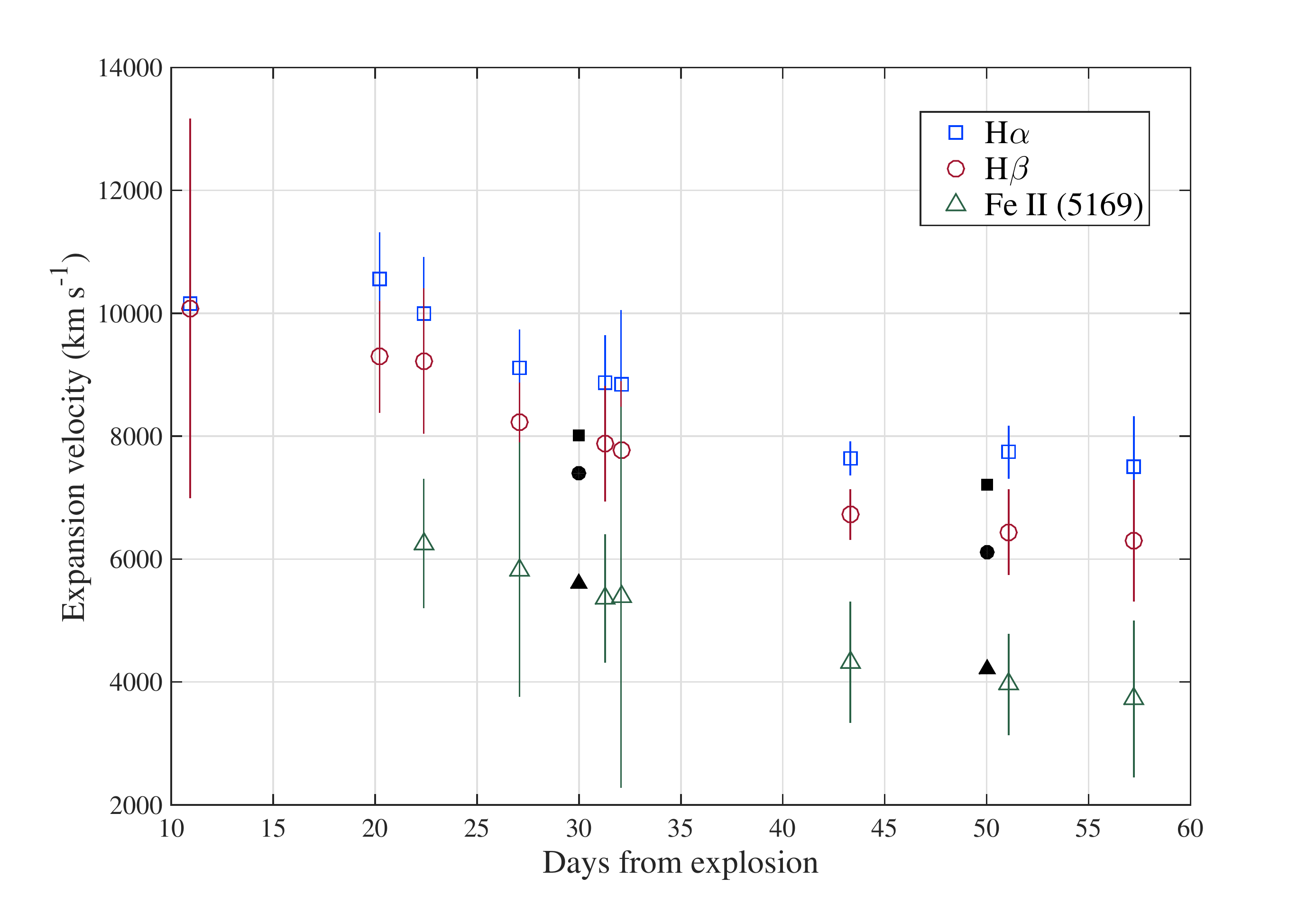}
\caption{Evolution of the expansion velocity of the SN ejecta for selected lines, between days 10 and 57 after explosion.
The derived expansion velocities (and the uncertainties) were obtained by fitting a parabola to the minima of
the P-Cygni absorption features.
The H$\alpha$ and H$\beta$ lines evolve from velocities around 10,000 to $\sim7000\ {\rm km\,s^{-1}}$
during this time interval, whereas the velocity of the Fe~II $\lambda5169$ line, visible from day 22 onward, 
decreases from around $6000$ to $\ltorder4000\ {\rm km\,s^{-1}}$ by day 57. 
Overplotted with filled black markers are the expansion velocities (for the three lines:  H$\alpha$, H$\beta$, and an average of Fe~II lines)
of the standard Type II-P SN\,2004et at days 30 and 50, as presented by ref.~19. 
The obtained trend and values of the expansion velocities are in broad agreement with typical SNe~II-P\cite{2006ApJ...645..841N, 2014MNRAS.442..844F}.
\label{SIfig-velevol}}
\end{SIfigure}

\begin{SIfigure}
\centering
\includegraphics[width=18cm]{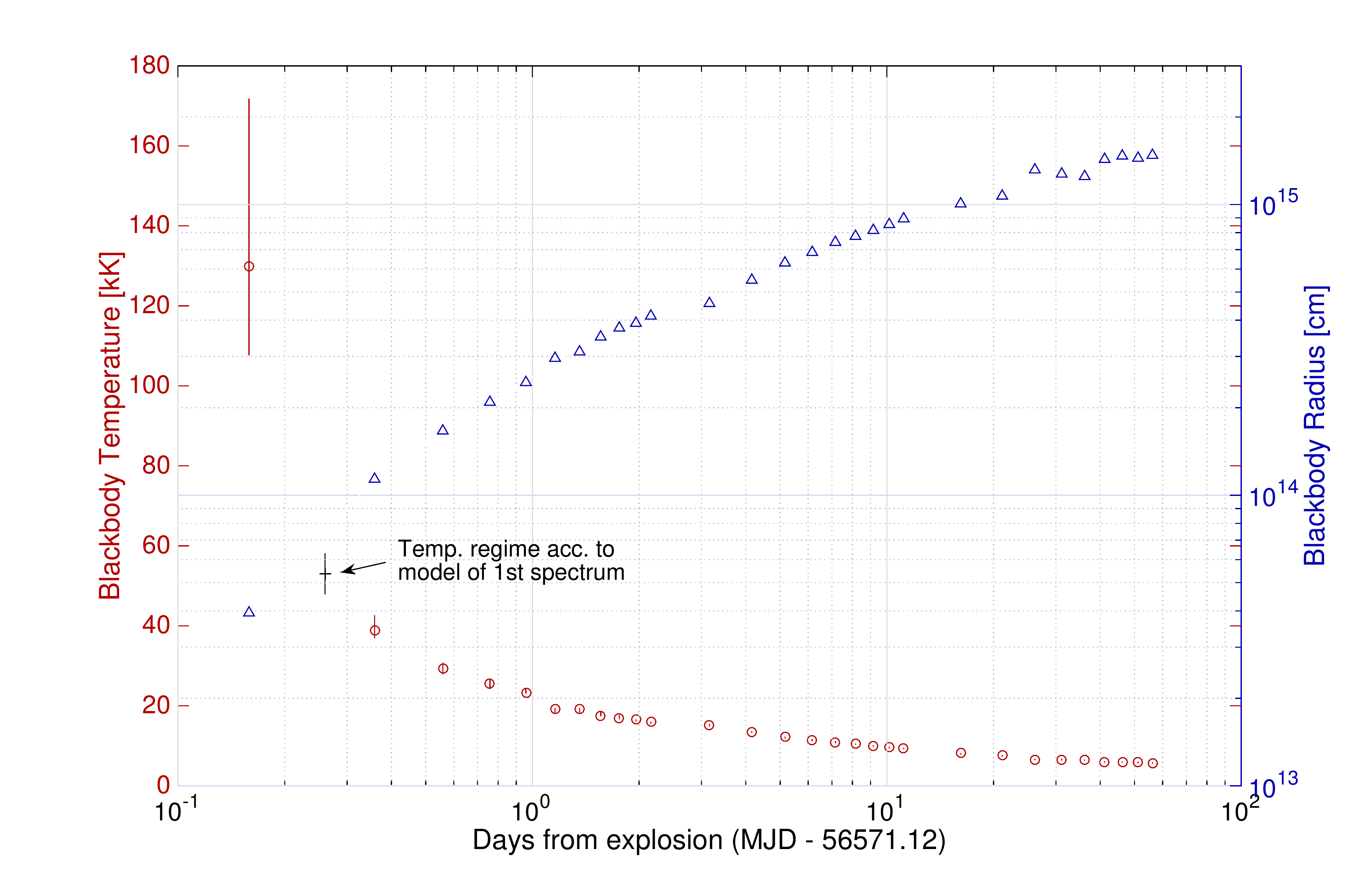}
\caption{Evolution of the estimated BB temperature and radius over the first 60\,d after explosion, based on the multiband photometry measurements (Fig.~\ref{fig-photometry}).
The first BB temperature estimate was obtained via careful extrapolation of the UVOT $UVM2$ and P60 $g+i$ light curves back to the first P48 (detection) point (see inset of Fig.~\ref{fig-photometry} and text for details). The early-time BB temperature estimates, within the first half day after explosion, are also in agreement with our temperature estimates from the modeling of the early Keck spectra (Fig.~\ref{fig-Groh_spec}), showing the highly ionised emission lines at temperatures $\gtorder50$\,kK.
\label{SIfig-BBTempRad}}
\end{SIfigure}

\begin{SIfigure}
\centering
\includegraphics[width=16cm]{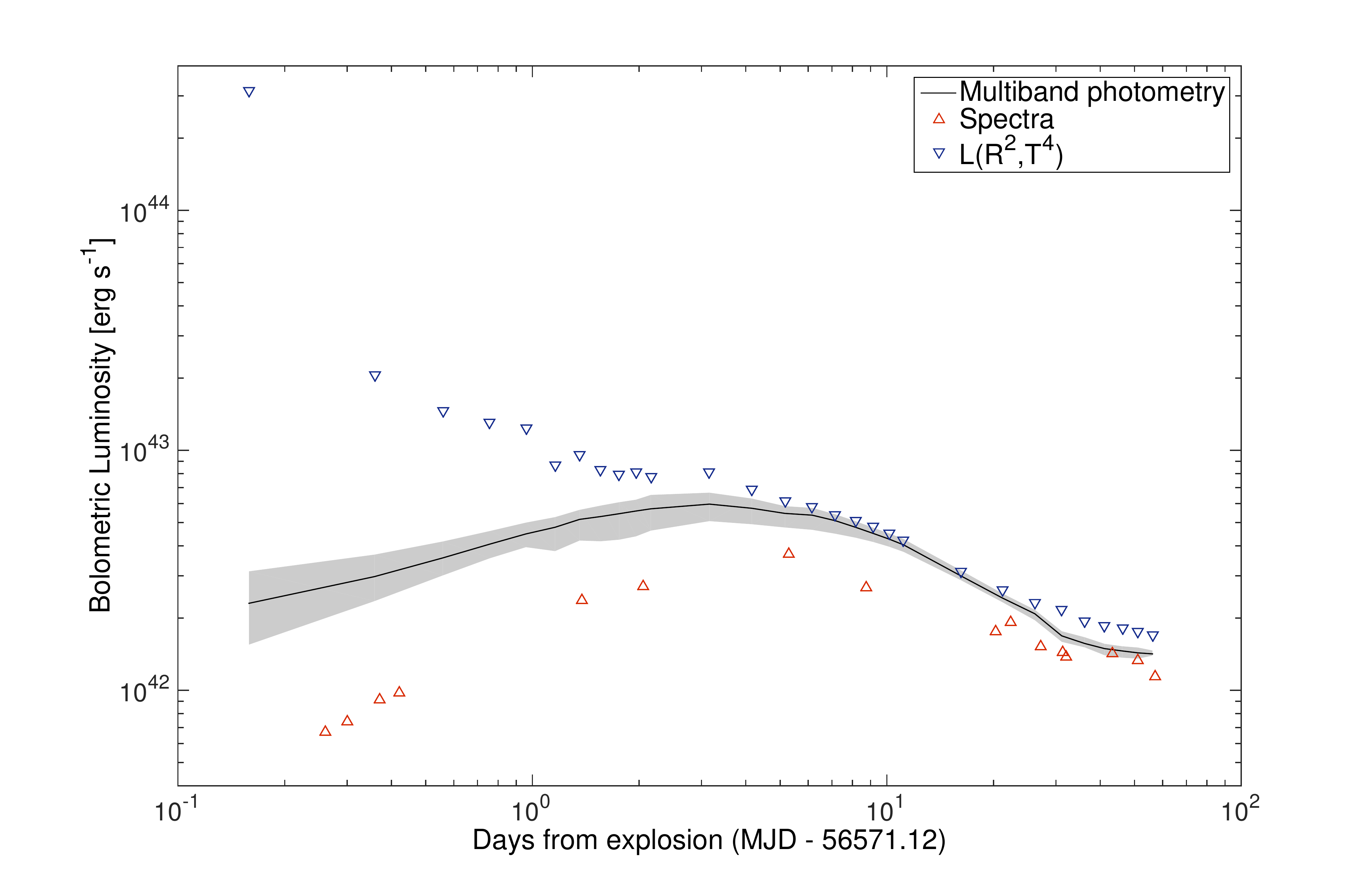}
\caption{Bolometric luminosity estimates over the first 60\,d after explosion.
The shaded region and the solid black line (a running mean of the region) denote bolometric luminosity estimates based on the multiband photometry (Fig.~\ref{fig-photometry}) according to three methods used to calculate the total flux from the SED: interpolation, order-4 polynomial fit, and BB fits. The top end of the shaded region can be regarded as our best lower limit on the real bolometric luminosity, based on the photometric observations.
The red triangles denote a (more conservative) lower limit on the bolometric luminosity obtained from our spectra (Fig.~\ref{fig-earlyspec}, Fig.~\ref{fig-latespec}), beginning with the early set of 4 Keck spectra at $\ltorder10$\,hr after explosion and ending with our latest spectrum at $57.2$ days.
The blue triangles show the luminosity as obtained by our best BB temperature and radius estimates (Fig.~\ref{SIfig-BBTempRad}), $L=4\pi R^2\sigma T^4$; the luminosity in the first point, at $\sim 3.8$\,hr after explosion, exceeds $10^{44} {\rm \,erg\,s^{-1}}$.
\label{SIfig-BolLum}}
\end{SIfigure}

\newpage

\begin{SIfigure}
\centering
\includegraphics[width=16cm]{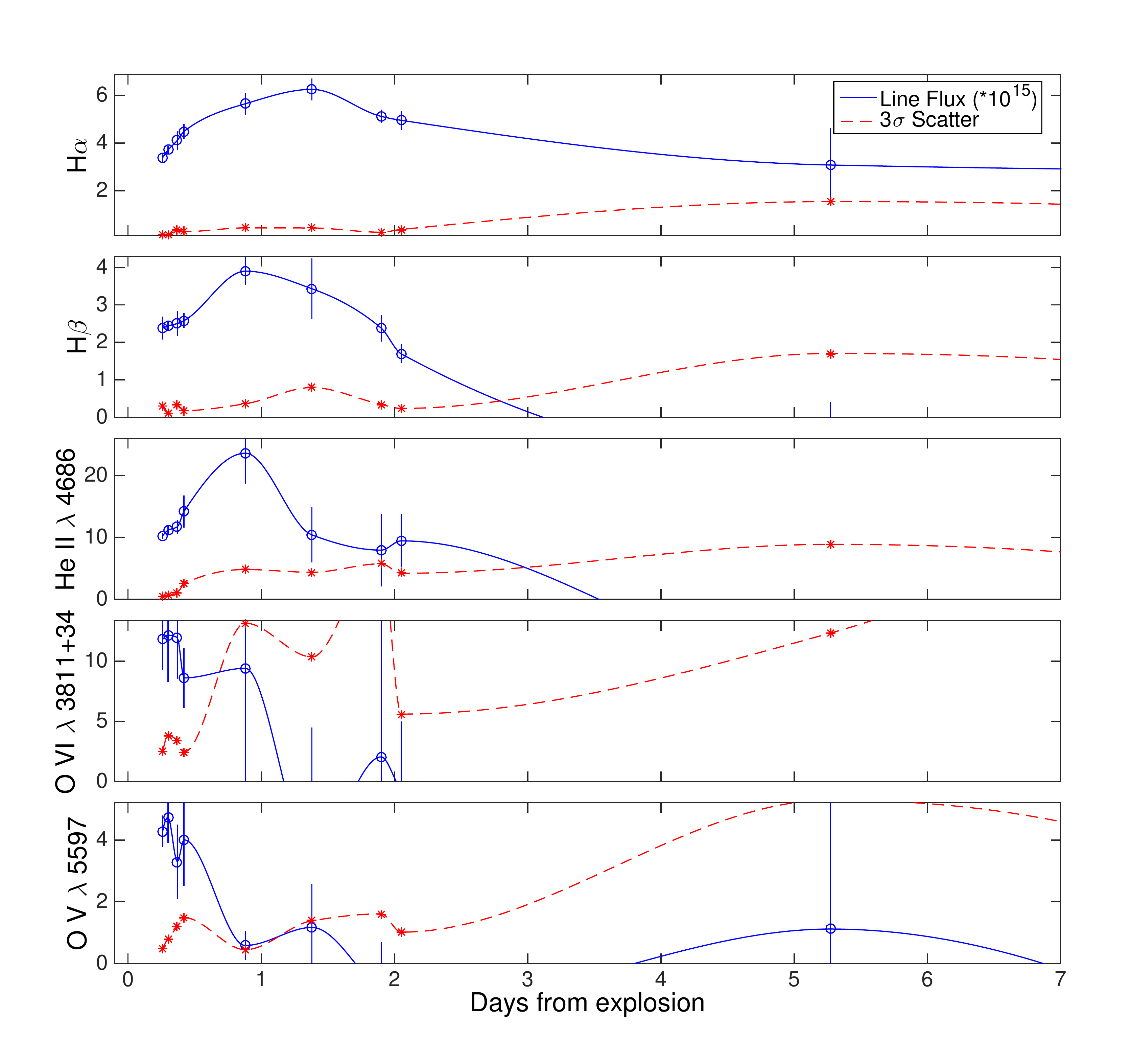}
\caption{Emission-line flux evolution during the first week after explosion.
A 3$\sigma$ uncertainty estimate based on the scatter is also plotted.
Note the general shape of the H$\alpha$, H$\beta$, and He~II emission-line flux curves, 
increasing and declining during the first $\sim2$\,d, whereas the highly ionised oxygen lines disappear completely 
(quickly falling to the background scatter) well within the first day after explosion. 
\label{SIfig-lineflux}}
\end{SIfigure}

\begin{SIfigure}
\centering
\includegraphics[width=11cm]{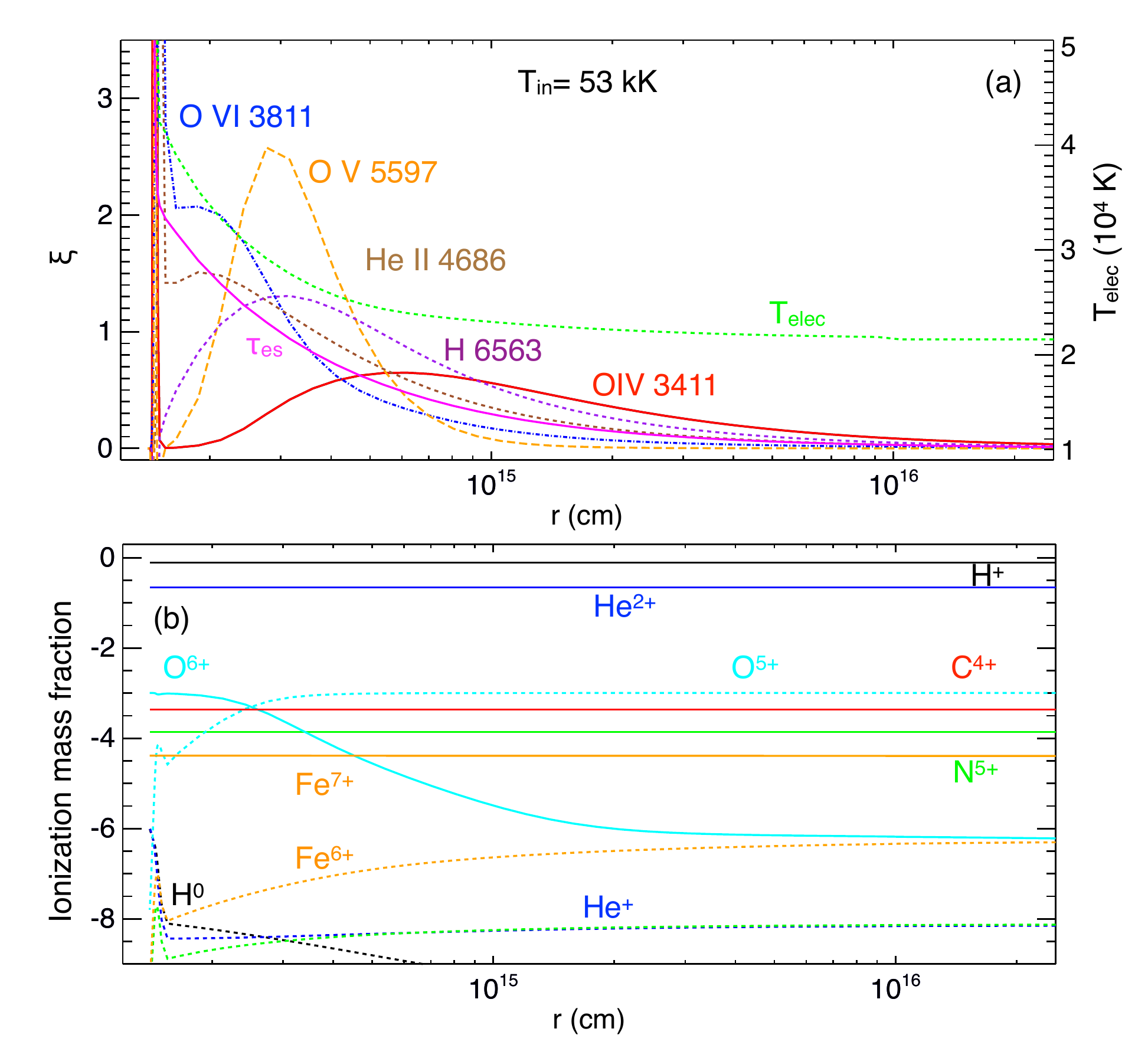}
\includegraphics[width=12cm]{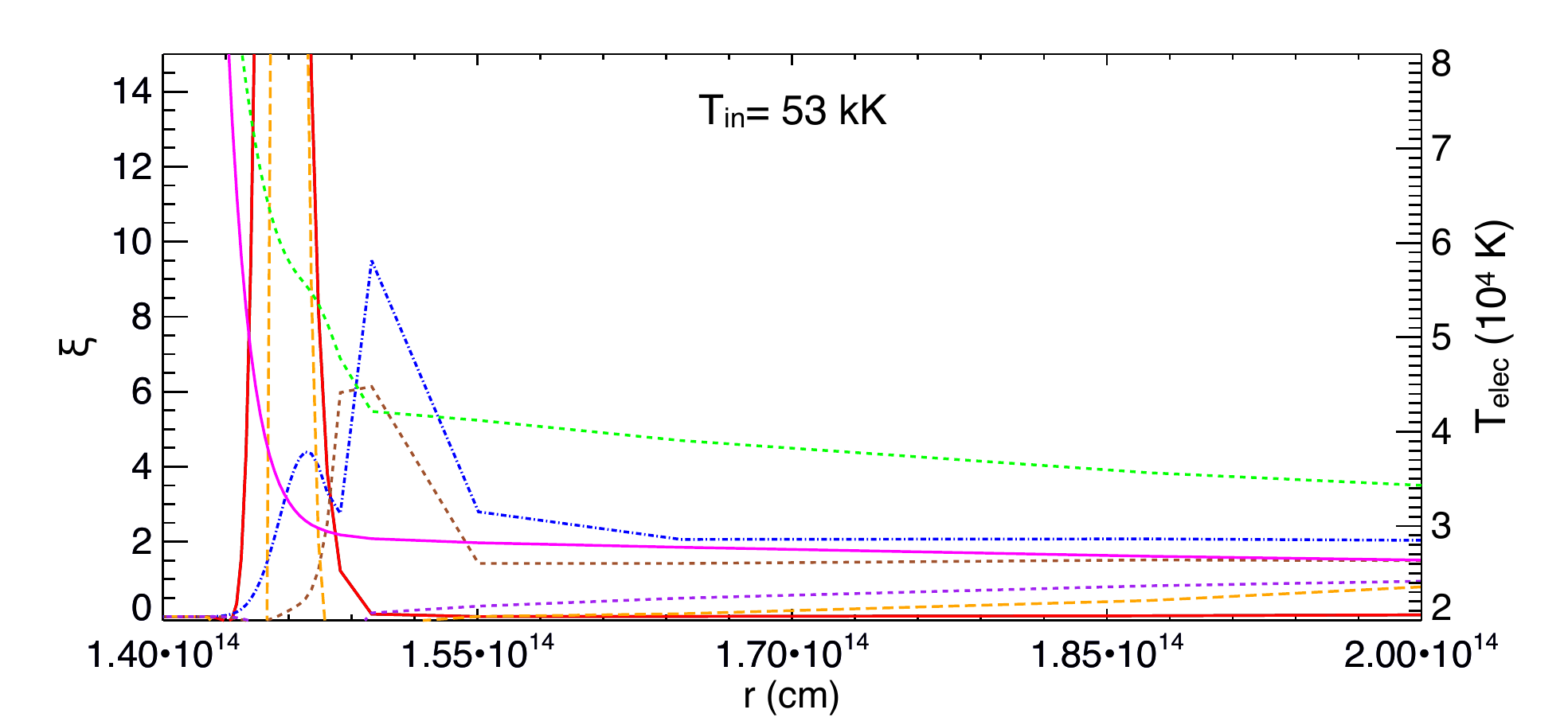}
\caption{Line-forming regions (top and bottom panels) and ionisation structure (middle) of the 53\,kK model, fitting the 6\,hr spectrum.
The quantity $\xi$ is related to the equivalent width of the line (following ref.~65) 
as EW $=\int_{R_{\rm in}}^{\infty} \xi(r)d({\rm log}\,r)$.
The bottom panel shows a close-up view of the very inner region, around   $R_{\rm in}$ (colour coding is identical to the top panel.) 
Radial profiles of the electron optical depth and electron temperatures are shown in the top panel.
\label{SIfig-Groh_line_form}}
\end{SIfigure}

\newpage

\begin{SIfigure}
\centering
\includegraphics[width=16cm]{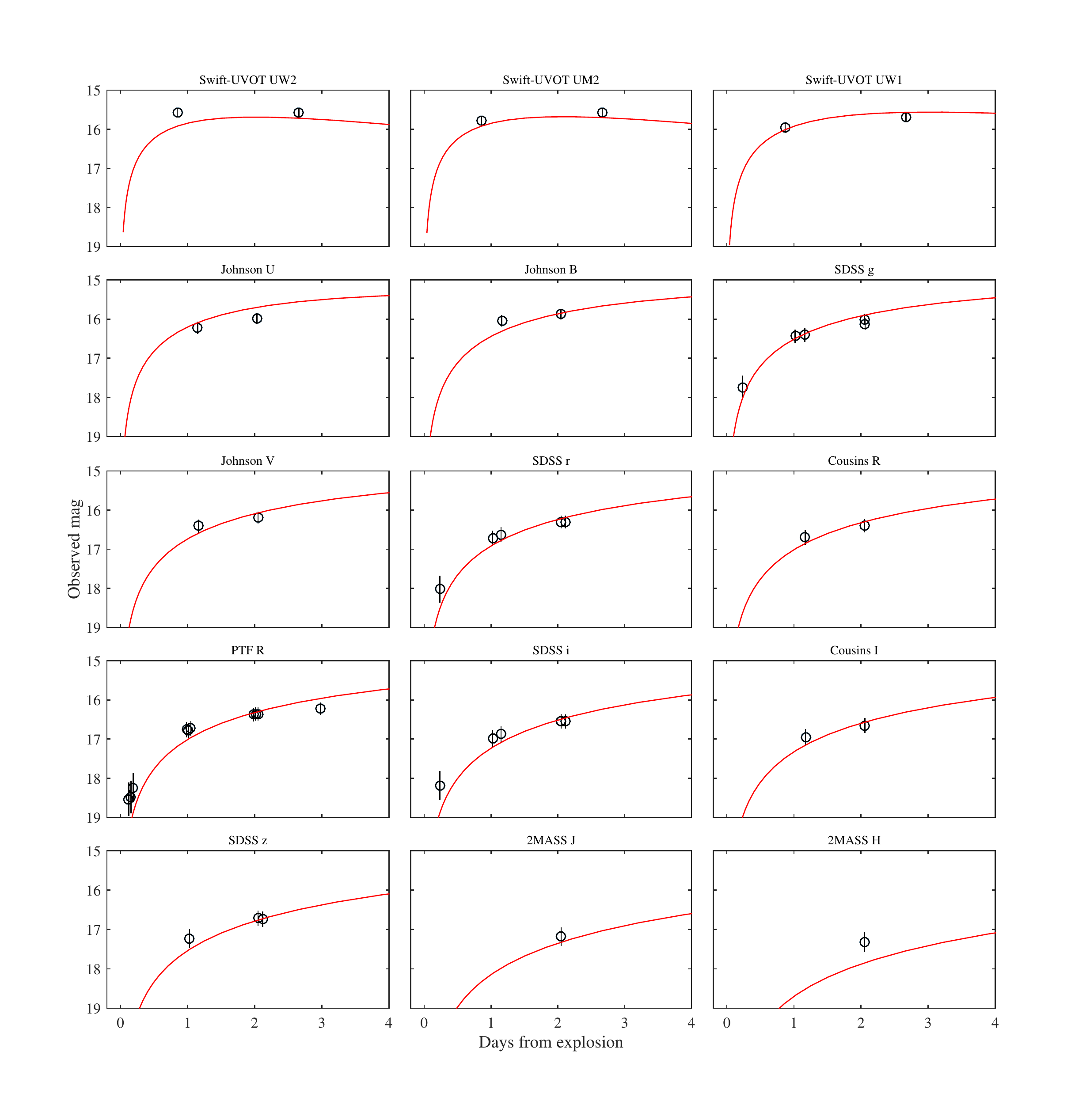}
\caption{Multicolour fits of RW11\cite{2011ApJ...728...63R} models to the early-time photometry.
The uncertainties have been scaled such that the minimal $\chi^2$ per degree of freedom equals 1 (and the original relative errors in between measurements are maintained). See ref.~25, 
presenting $R$-band fits of RW11 for a large sample of SNe~II from the PTF and iPTF surveys.
\label{SIfig-RW11_LCs}}
\end{SIfigure}

\begin{SIfigure}
\centering
\includegraphics[width=15cm]{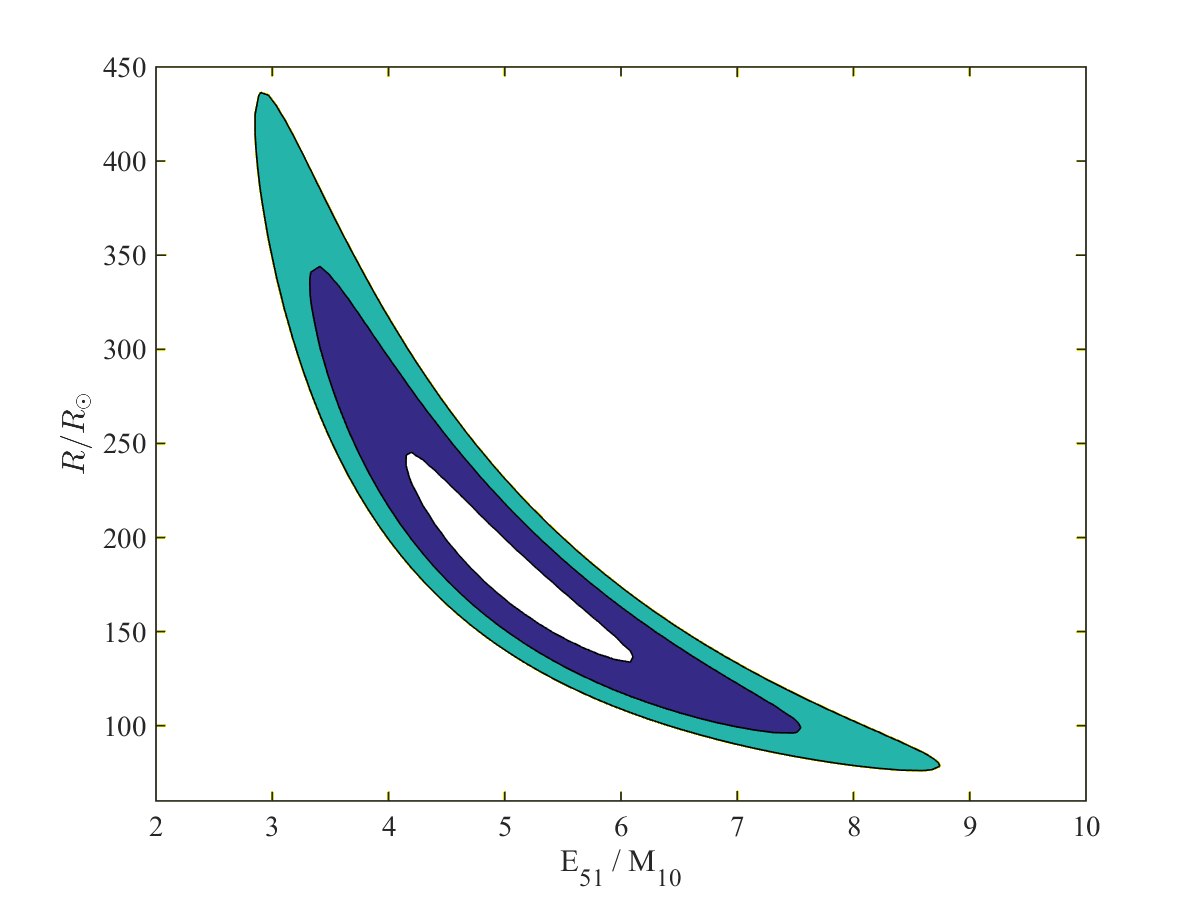}
\caption{Analysis of our multicolour UVOIR early-time data using the methods of ref.~24 
constrains the progenitor radius and explosion energy per unit mass. 
$1, 2, 3\sigma$ contours are plotted, as derived from $\chi^2$ fits to all the data (see ref.~25 
for details).
\label{SIfig-chi_RW11}}
\end{SIfigure}


\begin{SItable}\footnotesize
\begin{tabular}{llllll}
\hline\hline
UTC Obs-Date & Obs-Time & MJD & Phase$^a$ & Telescope/Instrument & Observer/Reducer \\
\hline
2013-10-06	& 09:10:36	& 56571.3824 & 6.2\,hr	& Keck-I/LRIS			& D. Perley / D. Perley \\
2013-10-06 	& 10:00:46	& 56571.4172 & 7.2\,hr	& Keck-I/LRIS			& D. Perley / D. Perley \\
2013-10-06 	& 11:39:05	& 56571.4855 & 8.9\,hr	& Keck-I/LRIS			& D. Perley / D. Perley \\
2013-10-06 	& 13:04:19	& 56571.5447 & 10.1\,hr	& Keck-I/LRIS			& D. Perley / D. Perley \\
2013-10-06 	& 13:13:49	& 56571.5513 & 10.3\,hr	& Keck-I/LRIS (Hi-Res)	& D. Perley / D. Perley \\
2013-10-07       & 00:02:18	& 56572.0016 & 21.1\,hr	& NOT/ALFOSC		& N. E. Groeneboom / F. Taddia \\
2013-10-07 	& 12:06:39	& 56572.5046 & 1.4\,d	& FTS/FLOYDS		& D. Sand / S. Valenti \\
2013-10-08	& 00:21:53	& 56573.0152 & 1.9\,d	& NOT/ALFOSC		& N. E. Groeneboom / F. Taddia \\
2013-10-08	& 04:00:46	& 56573.1672 & 2.0\,d	& P200/DBSP			& Y. Cao / Y. Cao\\
2013-10-08	& 06:34:07	& 56573.2737 & 2.1\,d	& Keck-II/DEIMOS		& K. Clubb, M. Graham / P. Kelly \\
2013-10-11	& 04:48:12	& 56576.2001 & 5.1\,d	& Keck-II/DEIMOS		& S. Tang / Y. Cao \\
2013-10-11	& 09:19:19	& 56576.3884 & 5.3\,d	& FTS/FLOYDS		& D. Sand / I. Arcavi, S. Valenti\\
\hline
2013-10-14 	& 21:05:02	& 56579.8785 & 8.8\,d	& WHT/ISIS			& WHT service / K. Maguire\\
2013-10-17	& 01:10:33	& 56582.0497 & 10.9\,d	& NOT/ALFOSC		& A. A. Djupvik / F. Taddia\\
2013-10-26	& 08:06:13	& 56591.3377 & 20.2\,d	& Lick-3\,m/Kast			& I. Shivvers, J. C. Mauerhan / S. B. Cenko\\
2013-10-28	& 12:05:01	& 56593.5035 & 22.4\,d	& FTS/FLOYDS		& S. Valenti / S. Valenti \\
2013-11-02	& 04:59:51	& 56598.2082 & 27.1\,d	& P200/DBSP			& A. Waszczak / P. Vreeswijk\\
2013-11-06	& 09:36:49	& 56602.4006 & 31.3\,d	& FTS/FLOYDS		& S. Valenti / S. Valenti \\
2013-11-07	& 04:25:03	& 56603.1841 & 32.0\,d	& APO/DIS			& M. Kasliwal / Y. Cao \\
2013-11-18	& 10:12:19	& 56614.4252 & 43.3\,d	& FTS/FLOYDS		& S. Valenti / S. Valenti \\
2013-11-26	& 04:50:56	& 56622.2020 & 51.1\,d	& P200/DBSP			& A. Waszczak / O. Yaron \\
2013-12-02	& 08:15:32	& 56628.3441 & 57.2\,d	& Keck-I/LRIS			& Y. Cao, D. Perley / D. Perley \\
\hline\hline
\end{tabular}
\caption{Log of Spectroscopy. 
Additional meta-data and the full set of spectra are publicly available through WISeREP (http://wiserep.weizmann.ac.il).
Notes: $^a$Hours/days with respect to the estimated explosion time (at ${\rm MJD_0}=56571.12$).
\label{SItab-spec}}
\end{SItable}


\end{document}